\documentclass[aps,pre,showpacs,preprintnumbers,amsmath,amssymb,twocolumn,superscriptaddress]{revtex4-1}
\usepackage{graphicx}
\usepackage{dcolumn}
\usepackage{bm}
\usepackage{tabularx}
\usepackage{amssymb}
\usepackage{array}
\usepackage{amsmath} 
\usepackage{upgreek}

\usepackage{longtable}

\usepackage{verbatim}
\usepackage{color}
\usepackage{hyperref}

\newcommand{\transp}{^\mathrm{T}}
\newcommand{\mat}[1]{\mathbf{#1}}

\newcommand{\tmat}[1]{\tilde{\mathbf{#1}}}

\begin{document}

\title{Marginally compact fractal trees with semiflexibility}

\author{Maxim Dolgushev}
\email{dolgushev@physik.uni-freiburg.de}
\affiliation {Institute of Physics, University of Freiburg, Hermann-Herder-Strasse 3, D-79104 Freiburg, Germany}
\affiliation {Institut Charles Sadron, Universit\'e de Strasbourg \& CNRS, 23 rue du Loess, 67034 Strasbourg Cedex,
France}

\author{Adrian L. Hauber}
\author{Philipp Pelagejcev}
\affiliation {Institute of Physics, University of Freiburg, Hermann-Herder-Strasse 3, D-79104 Freiburg, Germany}

\author{Joachim P. Wittmer}
\affiliation {Institut Charles Sadron, Universit\'e de Strasbourg \& CNRS, 23 rue du Loess, 67034 Strasbourg Cedex,
France}

\begin{abstract}

We study marginally compact macromolecular trees that are created by means of two different fractal generators. In doing so, we assume Gaussian statistics for the vectors connecting nodes of the trees. Moreover, we introduce bond-bond correlations that make the trees locally semiflexible. The symmetry of the structures allows an iterative construction of full sets of eigenmodes (notwithstanding the additional interactions that are present due to semiflexibility constraints), enabling us to get physical insights about the trees' behavior and to consider larger structures. Due to the local stiffness the self-contact density gets drastically reduced.

\end{abstract}
\maketitle

\section{Introduction}\label{intro}

In nature, many objects can be successfully represented through fractal models \cite{mandelbrot82,fractalnat1,fractalnat2,fractalnat3,fractalnat4,reuveni08,reuveni10,lieberman09,mirny11,fudenberg11,grosberg12,fractalnat5}. Examples are provided by lungs \cite{fractalnat1,fractalnat2,fractalnat3}, plants \cite{fractalnat4}, proteins \cite{reuveni08,reuveni10}, and chromatin \cite{lieberman09,mirny11,fudenberg11,grosberg12,fractalnat5}, to name only a few. Also man-made materials, such as super-repellent surfaces \cite{fractalmat1,fractalmat2}, porous cements \cite{fractalmat3}, super-lenses \cite{fractalmat4}, and supercapacitors \cite{fractalmat5,fractalmat6}, can be build in a fractal way in order to make a better performance. The purpose of many of these examples requests an effective usage of the space provided for them. This challenge is usually connected to a very dense packing of the objects \cite{lieberman09,nechaev17} and at the same time to a huge surface needed for their function (e.g., surface available for charge in case of supercapacitors \cite{fractalmat6}). Thus, in the best case almost all their constituents build a surface, e.g., for compact objects in three dimensional space consisting of $N$ units and having size $R\sim N^{1/3}$, the surface $A$ scales as $A\sim R^3\sim N$ \cite{dolgushev17a}. 

\begin{figure}[t]
\centering
\includegraphics[width=1\linewidth,trim=0 0 0 0]{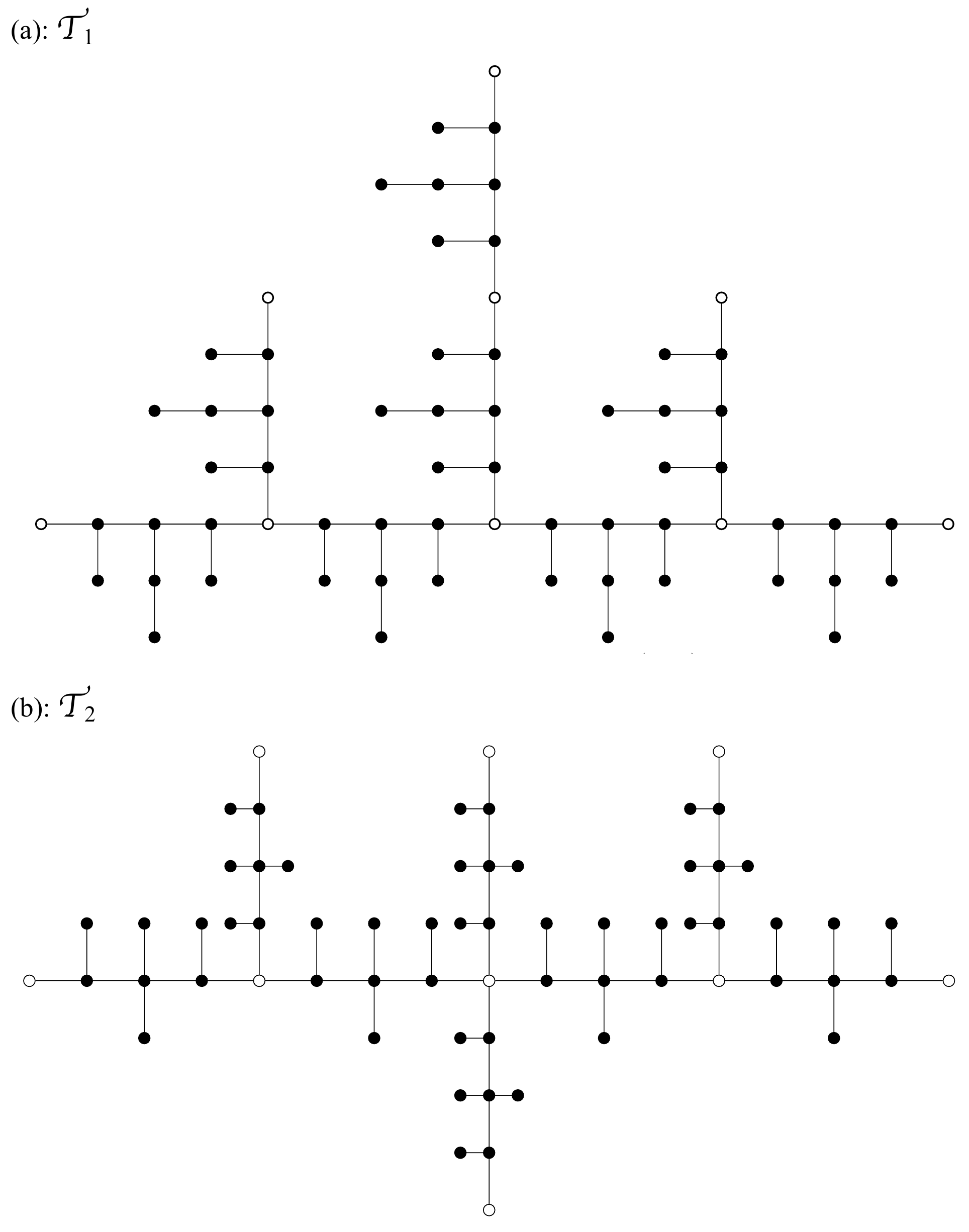}
\caption{Fractal trees $\mathcal{T}_1$ (a) and $\mathcal{T}_2$ (b) studied in this work. Both structures are at iteration $I=2$. Beads shown as open circles represent the trees of inital iteration $I=1$. The sketch is aimed only to present the topology of the fractal trees; their spatial conformations may appear in vastly different forms.}\label{trees}
\end{figure}

With respect to the biological and technological examples listed above, it is worth mentioning another actively studied system -- the melt of nonconcatenated and unknotted ring polymers \cite{muller96,muller00,halverson11a,halverson11b,obukhov14,grosberg14,sakaue11,rosa14,michieletto16,ge16,smrek16,kapnistos08,goossen14} -- that have been surmised to be marginally compact \cite{obukhov14,rosa14,smrek16}.
However, the marginal compactness of ring melts is controversially argued, partly due to the clever theoretical argument \cite{halverson11a} that the marginal compactness leads to a logarithmic divergence of the self-contact density. In a recent work \cite{dolgushev17a} by some of us, it was suggested a practical way out of this difficulty. There we have studied the fractal trees of Ref. \cite{polinska14} (see tree $\mathcal{T}_1$ of Fig.~\ref{trees}) that are by construction marginally compact. These toy-structures, not aiming to describe the full complexity of examples such as given by Refs. \cite{fractalnat1,fractalnat2,fractalnat3,fractalnat4,reuveni08,reuveni10,lieberman09,mirny11,fudenberg11,grosberg12,fractalnat5,fractalmat1,fractalmat2,fractalmat3,fractalmat4,fractalmat5,fractalmat6,muller96,muller00,halverson11a,halverson11b,obukhov14,grosberg14,sakaue11,rosa14,michieletto16,ge16,smrek16,kapnistos08,goossen14},  allowed us to show that a simple ingredient that can suppress the divergent behavior of the self-contact density $\hat{\rho}_c$ is the linear spacers between branching points of the trees. 

The present study focuses on another aspect of marginally compact trees, namely on the role of local semiflexibility. The recent studies \cite{polinska14,dolgushev17a} have considered Gaussian, marginally compact trees with interactions between topologically nearest neighboring beads, i.e. in the framework of generalized Rouse model \cite{gurtovenko05}. In particular, this assumption implies that the orientations of bonds are uncorrelated \cite{gurtovenko05,doi88}. However, the price one has to pay for the bond-correlations is a more complex structure of the dynamical matrix, that then in the easiest case (under freely-rotating bonds assumption for the non-adjacent bonds \footnote{We note that in the framework used here (which stems from Ref.~\cite{bixon78} for linear chains) the length of each bond is not constant (unlike for the classical freely-rotating chain model). Each bond can fluctuate with the fixed mean-square length $b^2$. This assumption allows to describe polymers by a (multivariate) Gaussian distribution.}) contains also the elements related to the next-nearest neighboring beads~\cite{dolgushev09a}. Notwithstanding this difficulty, the framework of semiflexible treelike polymers (STP) of Ref.~\cite{dolgushev09a}, where the semiflexibility is introduced at all beads (also at branching nodes), turned out to be very helpful in studying the relaxation dynamics of semiflexible dendrimers~\cite{fuerstenberg12,grimm16} and fractals~\cite{fuerstenberg13,mielke16}. Moreover, inclusion of bond-bond correlations has been shown to have a fundamental importance for NMR relaxation of dendrimers \cite{kumar13b,markelov15,markelov16,shavykin16}. Therefore, the semiflexiblity should also be an important ingredient for marginally compact trees.

In this work we consider marginally compact trees which are locally semiflexible. The topology of the trees is sketched in Fig.~\ref{trees}. Fractal tree $\mathcal{T}_1$ consists of beads of functionality $1$, $2$, and $3$; the generalized Rouse \cite{gurtovenko05} behavior (i.e. in the absence of bond-bond correlations) of these trees has been studied in Refs.~\cite{polinska14,dolgushev17a}. In order to make our results more rigorous and to exemplify the role of functionality of branching nodes we introduce another fractal generator that builds marginally compact trees $\mathcal{T}_2$ (see Fig.~\ref{trees}), which do not have any linear spacers but contain beads of functionality $4$. Both trees $\mathcal{T}_1$ and $\mathcal{T}_2$ show all relevant scalings of marginally compact, flexible trees \cite{polinska14,dolgushev17a}, when one introduces local bending rigidity. At the same time the semiflexibility leads to a swelling of the structures and hence to an increase of the higher relaxation times and to a significant suppression of self-contacts. Yet, the underlying STP framework~\cite{dolgushev09a} allows us to perform a detailed analysis of eigenmodes and to reduce the computational work.  

The paper is structured as follows. In the next section we provide theoretical formulas and details for the dynamical matrix in the STP framework~\cite{dolgushev09a}, whose spectra for trees $\mathcal{T}_1$ and $\mathcal{T}_2$ are analyzed then in Sec.~\ref{spectrum} (the technical details are relegated to the Appendix). The static and dynamical properties of the trees are presented in Sec.~\ref{properties}. Section~\ref{conclusions} closes the paper with a summary and conclusions.   

\section{Theoretical model}\label{model} 

We start this section with a brief recall of the theory of semiflexible treelike polymers (STP) \cite{dolgushev09a}. The STP framework allows to introduce local bending rigidity for Gaussian trees with arbitrary topology. The resulting dynamical matrix of the trees is sparse and has an analytically closed form.

In the STP theory the edges of the treelike structures represent Gaussian bonds $\{\mathbf{d}_i\}$, whose orientations are constrained. For any two adjacent bonds $\mathbf{d}_i$ and $\mathbf{d}_j$ one has $\langle \mathbf{d}_i \cdot \mathbf{d}_j\rangle=\pm b^2q_m$, where $b^2=\langle \mathbf{d}_i \cdot \mathbf{d}_i\rangle=\langle \mathbf{d}_j \cdot \mathbf{d}_j\rangle$ is the mean-square length of each bond and $q_m$ is the so-called stiffness parameter related to bead $m$ connecting bonds $\mathbf{d}_i$ and $\mathbf{d}_j$. The sign determines connection of the bonds, plus sign corresponds to a head-to-tail connection and minus to two other configurations. The connection between non-adjacent bonds is taken in a freely-rotating manner, i.e., for bonds connected through the path $\mathbf{d}_{k_1},...,\mathbf{d}_{k_s}$ the relation $ \langle \mathbf{d}_i\cdot \mathbf{d}_j \rangle=\langle \mathbf{d}_i\cdot \mathbf{d}_{k_1}\rangle {\langle \mathbf{d}_{k_1}\cdot\mathbf{d}_{k_2}\rangle}\cdots\langle\mathbf{d}_{k_s}\cdot \mathbf{d}_j\rangle b^{-2s}$ holds \cite{Note1}.

Given that each bond $\mathbf{d}_i$ has a zero mean, the average scalar products $\{\langle \mathbf{d}_i \cdot \mathbf{d}_j\rangle\}$ represent the covariance matrix ${\bf\Sigma}=(\langle \mathbf{d}_i \cdot \mathbf{d}_j\rangle)$ that fully determines the Gaussian distribution of the bonds. Furthermore, each bond vector $\mathbf{d}_i$ can be represented through a difference of position vectors of beads connected through $\mathbf{d}_i$, $\mathbf{d}_i=\mathbf{r}_n-\mathbf{r}_m$. With this, the potential energy of the tree, 
\begin{equation}
V=\frac{3}{2}k_BT\sum_{i,j}({\bf\Sigma}^{-1})_{ij}\mathbf{d}_i \cdot \mathbf{d}_j=\frac{3k_BT}{2b^2}\sum_{m,n}A_{nm}\mathbf{r}_n\cdot\mathbf{r}_m,
\end{equation}
is fully represented by the dynamical matrix $\mathbf{A}=(A_{nm})$. Based on the potential energy $V$, the dynamics of a polymer can be described by a set of Langevin equations, e.g., for the position of the $k$th bead one has
 \begin{align}
 \zeta \frac{\partial }{\partial t}\mathbf{r}_k(t)+\frac{3k_BT}{b^2}\sum_{n}A_{kn}\mathbf{r}_n=\mathbf{{g}}_k(t),
 \label{langevin}
 \end{align}
 where $\zeta \frac{\partial }{\partial t}\mathbf{r}_k(t)$ and $\mathbf{{g}}_k(t)$  are the friction and stochastic (white-noise) forces, respectively.

The conditions on the averaged scalar products used in the STP framework lead to an analytic form of $\mathbf{A}$. Moreover, under these conditions the matrix $\mathbf{A}$ turns out to be very sparse. Its non-vanishing elements are either diagonal or related to nearest-neighboring and next-nearest neighboring beads. For a bead of functionality $f$ (i.e. it has $f$ nearest neighbors) directly connected to beads of functionalities $f_1,\dots,f_f$ the diagonal element of $\mathbf{A}$ reads
\begin{equation}\label{Adiag}
\mu^{(f)}_{f_1\dots f_f}=\frac{f}{1-(f-1)q_f}+\sum_{s=1}^{f}\frac{(f_s-1)q_{f_s}^2}{1-(f_s-2)q_{f_s}-(f_s-1)q_{f_s}^2}.
\end{equation}
For two directly connected beads of functionalities $f_1$ and $f_2$ one has
\begin{equation}\label{ANN}
\nu_{f_1f_2}=-\frac{1-(f_1-1)(f_2-1)q_{f_1}q_{f_2}}{(1-(f_1-1)q_{f_1})(1-(f_2-1)q_{f_2})}
\end{equation}
and for two next-nearest neighboring beads connected through a bead of functionality $f$ the corresponding element of $\mathbf{A}$ is
\begin{equation}\label{ANNN}
\rho_{f}=\frac{q_f}{1-(f-2)q_f-(f-1)q_f^2}.
\end{equation}
In Eqs.~\eqref{Adiag}-\eqref{ANNN} the stiffness parameters $q_{f_i}$ are related to the beads (junctions) of functionality $f_i$. Each stiffness parameter $q_{f_i}$ is bounded from above by $1/(f_i-1)$ \cite{mansfield80,dolgushev09a}; if all stiffness parameters are zero one recovers fully-flexible structures so that the dynamical matrix $\mathbf{A}$ transforms into the connectivity (Laplacian) matrix. 

We note that the STP theory allows to choose the stiffness parameters at every junction separately. Here, however, a homogeneous case is used in which all junctions of the same functionality, say $f\geq2$, have the same stiffness parameter $q_f$. Moreover, here we assume a linear dependence of the stiffness parameters from each other by taking $q_f=q/(f-1)$ (with $f\geq2$), so that the limits $0$ and $1/(f-1)$ are reached simultaneously for all junctions by varying $q$ from $0$ to $1$.  For beads of functionality $1$ no stiffness parameter can be assigned. This fact is automatically taken by Eqs.~\eqref{Adiag}-\eqref{ANN} into account, where the corresponding terms due to prefactors like $(f_i-1)$ disappear.

Needless to say, the information about the behavior of STP in completely encoded in the eigenvalues and eigenvectors of the dynamical matrix $\mathbf{A}$.  Moreover, the symmetry of the structures allows to reduce computational efforts and to get physical insights of the relaxation behavior, as we proceed to show in Sec.~\ref{spectrum}. 

\section{Spectrum of the dynamical matrix and the corresponding eigenmodes}\label{spectrum} 

The symmetry of trees $\mathcal{T}_1$ and $\mathcal{T}_2$ allows an iterative construction of a full set of eigenvectors \footnote{The eigenvectors of the set are linearly independent, but not necessarily orthogonal.}. The construction procedure is rooted in the work of Cai and Chen~\cite{cai97} for flexible dendrimers, which has been extended to STP treatment of semiflexible dendrimers~\cite{fuerstenberg12,grimm16} and regular fractals~\cite{fuerstenberg13,mielke16}.

\begin{figure}[t]
\centering
\includegraphics[width=1\linewidth,trim=0 0 0 0]{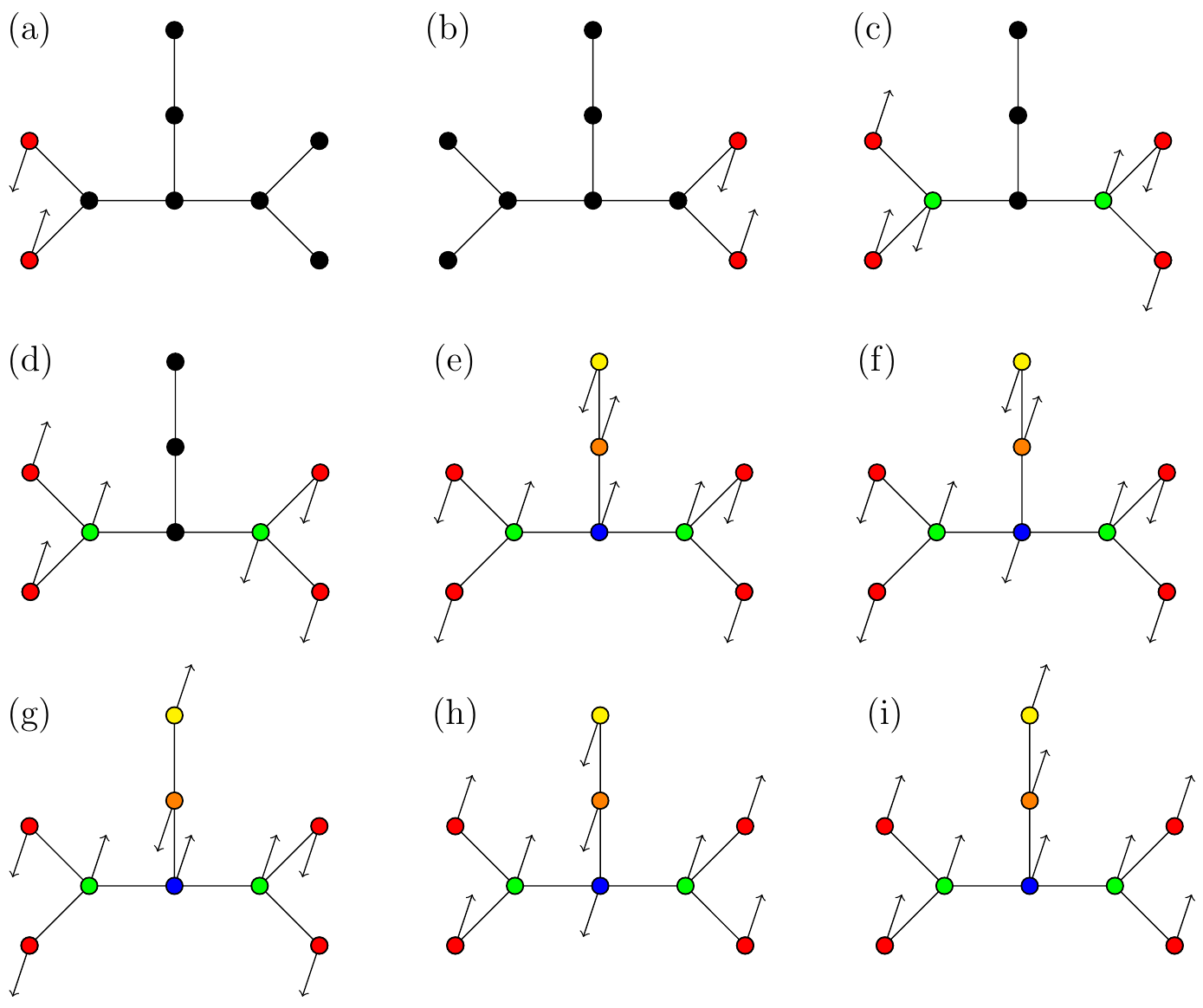}
\caption{Schematic sketch of eigenmodes of $\mathcal{T}_1$ for $I=1$. Beads having the same amplitude are color-coded. Black beads are immobile.}\label{modes_a}
\end{figure}

We start with tree $\mathcal{T}_1$ at iteration $I=1$. Figure~\ref{modes_a} displays the eigenmodes of the structure. Those of Fig.~\ref{modes_a}(a)-(d) leave some beads immobile, whereas in the eigenmodes of Fig.~\ref{modes_a}(e)-(i) all beads are involved. The modes (a) and (b) represent two vectors, which contain only two non-zero entries $1/\sqrt{2}$ and $-1/\sqrt{2}$. The ensuing (double degenerate) eigenvalue is equal to $\mu_3^{(1)}-\rho_3$, i.e. the $1\times1$ matrix describing this motion
\begin{align}
\mat{A}^{(1)}(\mathcal{T}_1)=
\begin{pmatrix}
\mu_3^{(1)}-\rho_3
\end{pmatrix}.\label{A1}
\end{align}
Next, we consider the modes displayed in Fig.~\ref{modes_a}(c)-(d) that have the shape $(x,x,\mp y,0,0,0,\pm y,-x,-x)^{\intercal}$. Multiplying the dynamical matrix with these vectors leads to a set of two non-trivial linear equations on $x$ and $y$ represented through the matrix 
\begin{align}
\tmat{A}^{(1)}(\mathcal{T}_1)=
\begin{pmatrix}
\mu_3^{(1)}+\rho_3 && \nu_{13}\\
2\nu_{13} && \mu_{113}^{(3)}-\rho_3
\end{pmatrix}.\label{At1}
\end{align}
Thus, diagonalization of $\tmat{A}^{(1)}(\mathcal{T}_1)$ given by Eq. \eqref{At1} leads to two eigenvalues of $\mathbf{A}$; the smallest one is related to Fig.~\ref{modes_a}(d) and the other one to Fig.~\ref{modes_a}(c). The remaining five eigenvalues of $\mathbf{A}$ are obtained from the diagonalization of the reduced matrix
\begin{align}
\mat{B}^{(1)}(\mathcal{T}_1)=\begin{pmatrix}
\mu_3^{(1)}+\rho_3 && \nu_{13} && 0 && 0 && \rho_3 \\
2\nu_{13} && \mu_{113}^{(3)}+\rho_3 && 0 && \rho_3 && \nu_{33}\\
0 && 0 && \mu_2^{(1)} && \nu_{12} && \rho_2 \\
0 && 2\rho_3 && \nu_{12} && \mu_{13}^{(2)} && \nu_{23}\\
4\rho_3 && 2\nu_{33} && \rho_2 && \nu_{23} && \mu_{233}^{(3)}
\end{pmatrix}.\label{B1}
\end{align}
This matrix is related to the eigenmodes of Fig.~\ref{modes_a}(e)-(i). For each of these modes the beads that are symmetric with respect to the core (blue bead) move in the same direction and with the same amplitude. Figure~\ref{modes_a}(i) depicts the translational mode $(1,\dots,1)^{\intercal}/\sqrt{N}=(1/3,\dots,1/3)^{\intercal}$ related to the eigenvalue $\lambda_0=0$.

\begin{figure}[t]
\centering
\includegraphics[width=1\linewidth,trim=0 0 0 0]{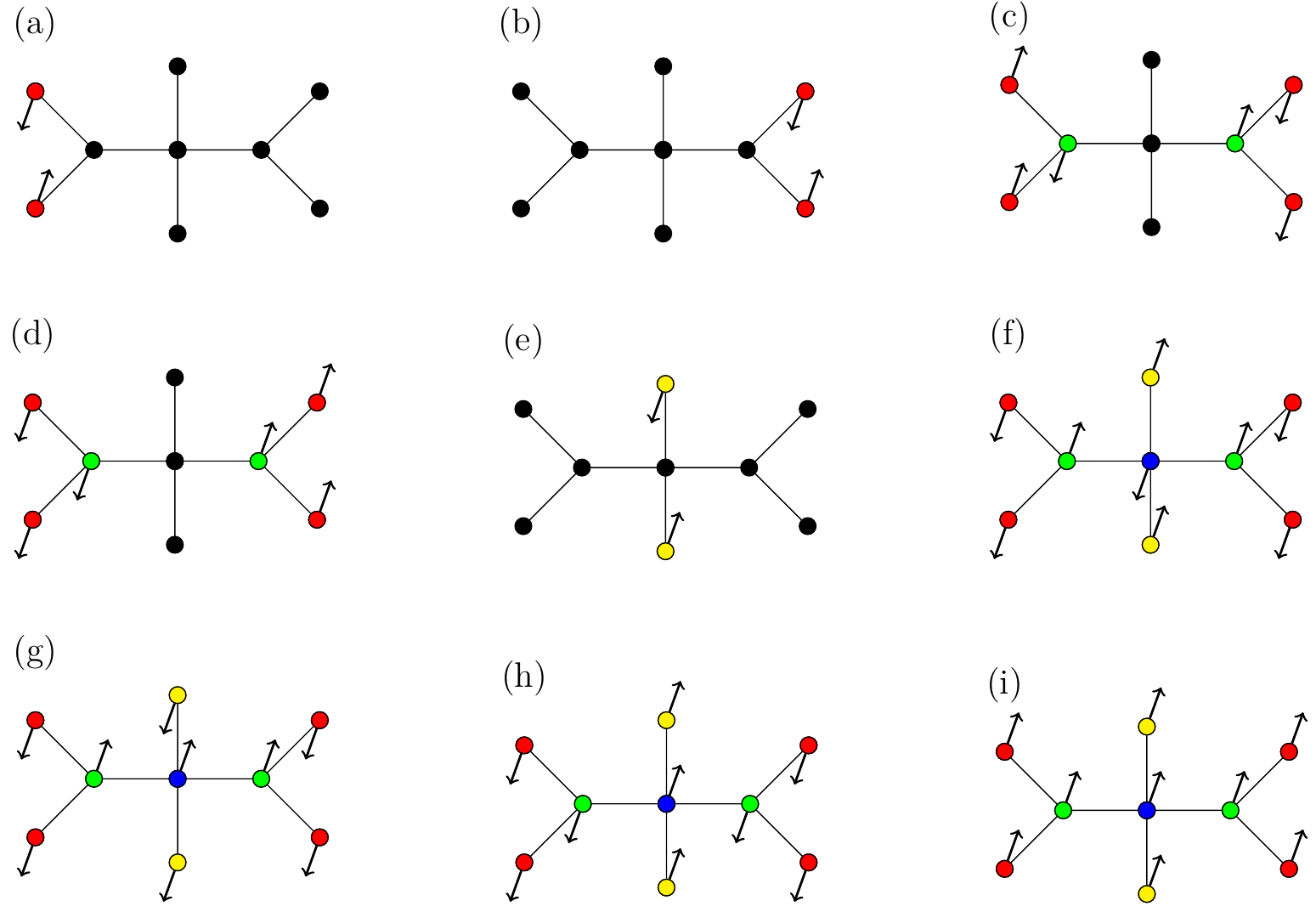}
\caption{Schematic sketch of eigenmodes of $\mathcal{T}_2$ for $I=1$. Beads having the same amplitude are color-coded. Black beads are immobile.}\label{modes_p}
\end{figure}

The construction of eigenmodes for tree $\mathcal{T}_2$ goes in a similar manner, see Fig.~\ref{modes_p}. The modes of Fig.~\ref{modes_p}(a)-(b) are related to the reduced matrix 
\begin{equation}
\mat{A}^{(1)}(\mathcal{T}_2)=\begin{pmatrix}
\mu_3^{(1)}-\rho_3
\end{pmatrix}
\end{equation}
that is equal to $\mat{A}^{(1)}(\mathcal{T}_1)$ of Eq.~\eqref{A1}. The matrix $\tmat{A}^{(1)}(\mathcal{T}_2)$ corresponding to Fig.~\ref{modes_p}(c)-(d) differs slightly from $\tmat{A}^{(1)}(\mathcal{T}_1)$ of Eq.~\eqref{At1} due to the core bead of functionality $4$,
\begin{align}
\tmat{A}^{(1)}(\mathcal{T}_2)=
\begin{pmatrix}
 \mu^{(1)}_3 + \rho_3 & \nu_{13} \\
 2\nu_{13} &\mu^{(3)}_{114}-\rho_4
\end{pmatrix}.\label{At1p}
\end{align}
Differently from $\mathcal{T}_1$, tree  $\mathcal{T}_2$ has for $I=1$ five eigenmodes that leave some beads (incl. the core) immobile. So the mode of Fig.~\ref{modes_p}(e) leads to the eigenvalue $\mu_4^{(1)}-\rho_4$, that can be formulated as $1\times1$ matrix 
\begin{equation}
\mat{\hat A}^{(1)}(\mathcal{T}_2)=\begin{pmatrix}
\mu_4^{(1)}-\rho_4
\end{pmatrix}.
\end{equation} 
The remaining four eigenvalues related to Fig.~\ref{modes_p}(f)-(i) come from the diagonalization of 
\begin{align}
\mat{B}^{(1)}(\mathcal{T}_2) =
\begin{pmatrix}
\mu_{1133}^{(4)} & 2\nu_{14} & 4\rho_3 & 2\nu_{34} \\
\nu_{14} & \mu^{(1)}_{4}+\rho_4 & 0 & 2\rho_4 \\
\rho_3 & 0 & \mu^{(1)}_3+\rho_3 & \nu_{13} \\
\nu_{34} & 2\rho_4 & 2\nu_{13} & \mu^{(3)}_{114}+\rho_4
\end{pmatrix}.
\end{align}
As for $\mathcal{T}_1$, this matrix has one vanishing eigenvalue related to the translational mode of Fig.~\ref{modes_p}(i).

The above procedure of construction of the sets of eigenmodes can be extended for higher $I>1$. The respective reduced matrices can be build iteratively, see Appendix. Here we discuss the sizes of the reduced matrices and the degeneracy of the corresponding eigenvalues. 

As it is observed for $I=1$, the modes (a) and (b) of Figs.~\ref{modes_a} and \ref{modes_p} lead to a double degenerate eigenvalue $(\mu_3^{(1)}-\rho_3)$. Going to the next iteration each bond gets replaced through a tree of iteration $I=1$ (see Fig.~\ref{trees}), hence each bead of functionality $1$ at iteration $I=1$ leads to a pattern as displayed in Fig.~\ref{modes_a}(a) at iteration $I=2$. At iteration $I-1$ trees $\mathcal{T}_1$ and $\mathcal{T}_2$ have $(3\cdot8^{I-1}+11)/7$ and $(4\cdot8^{I-1}+10)/7$ beads with functionality $1$, respectively. Thus the degeneracy of eigenvalue $(\mu_3^{(1)}-\rho_3)$ at iteration $I$ is $(3\cdot8^{I-1}+11)/7$ for $\mathcal{T}_1$  and $(4\cdot8^{I-1}+10)/7$ for $\mathcal{T}_2$. For tree $\mathcal{T}_2$ each bond of the previous iteration will lead to the pattern of Fig.~\ref{modes_p}(e). Hence the degeneracy of eigenvalue $(\mu_4^{(1)}-\rho_4)$ at iteration $I$ is equal to the number of bonds in $\mathcal{T}_2$ at iteration $I-1$, i.e. to $8^{I-1}$.

Now, going from one iteration to the next ($I-1\rightarrow I$), two next-nearest neighboring beads both of functionality $1$ [such as in involved in the eigenmode of Fig.~\ref{modes_a}(a)] lead to two directly connected trees $\mathcal{T}_1$ or $\mathcal{T}_2$ of $I=1$ (called leaves in the following, see Appendix). These leaves are involved in the eigenmodes, where each bead of one leaf has an opposite amplitude to that of the symmetrically equivalent bead of the other leaf. Moreover, in these modes all symmetrically equivalent beads belonging to the same leaf have the same amplitude and phase. In general, these modes lead to reduced matrices $\mat{A}^{(n)}(\mathcal{T}_1)$ and $\mat{A}^{(n)}(\mathcal{T}_2)$ whose iterative construction for $n=2,\dots,I$ is discussed in the Appendix. The size of  matrices $\mat{A}^{(n)}$ is
\begin{equation}\label{Sn}
S(n)=\begin{cases}
\frac{\sqrt{13}-1}{6\sqrt{13}}(4+\sqrt{13})^n+\frac{\sqrt{13}+1}{6\sqrt{13}}(4-\sqrt{13})^n \text{ for } \mathcal{T}_1,\\
\frac{\sqrt{37}-1}{6\sqrt{37}}\left(\frac{7+\sqrt{37}}{2}\right)^n+\frac{\sqrt{37}+1}{6\sqrt{37}}\left(\frac{7-\sqrt{37}}{2}\right)^n \text{ for } \mathcal{T}_2.\\
\end{cases}
\end{equation}
Following the above discussion, the degeneracy of each eigenvalue stemming from $\mat{A}^{(n)}$ appearing for the trees at iteration $I\geq n$ is
\begin{equation}
D(n)=\begin{cases}
(3\cdot8^{I-n}+11)/7 \text{ for } \mathcal{T}_1,\\
(4\cdot8^{I-n}+10)/7 \text{ for } \mathcal{T}_2.\\
\end{cases}
\end{equation}
The size $\hat{S}(n)$ of $\mat{\hat A}^{(n)}(\mathcal{T}_2)$ is equal to $S(n)$ of $\mathcal{T}_2$,
\begin{equation}
\hat{S}(n)=\small{\frac{\sqrt{37}-1}{6\sqrt{37}}\left(\frac{7+\sqrt{37}}{2}\right)^n+\frac{\sqrt{37}+1}{6\sqrt{37}}\left(\frac{7-\sqrt{37}}{2}\right)^n},
\end{equation}
and the degeneracy of each ensuing eigenvalue at iteration $I\geq n$ is (\textit{vide supra})
\begin{equation}
\hat{D}(n)=8^{I-n}.
\end{equation}

Apart from matrices $\mat{A}^{(1)}(\mathcal{T}_1),\dots,\mat{A}^{(I)}(\mathcal{T}_1)$ for $\mathcal{T}_1$ or $\mat{A}^{(1)}(\mathcal{T}_2),\dots\mat{A}^{(I)}(\mathcal{T}_2)$ and $\mat{\hat A}^{(1)}(\mathcal{T}_2),\dots,\mat{\hat A}^{(I)}(\mathcal{T}_2)$ for $\mathcal{T}_2$, there appear for each tree (at iteration $I$) one matrix $\tmat{A}^{(I)}$ and one matrix $\mat{B}^{(I)}$. The size of $\tmat{A}^{(I)}$ is
\begin{equation}\label{tildeSm}
\widetilde{S}(I)=\begin{cases}
\frac{\sqrt{13}+2}{6\sqrt{13}}(4+\sqrt{13})^I+\frac{\sqrt{13}-2}{6\sqrt{13}}(4-\sqrt{13})^I \text{ for } \mathcal{T}_1,\\
\frac{\sqrt{37}+5}{6\sqrt{37}}\left(\frac{7+\sqrt{37}}{2}\right)^I+\frac{\sqrt{37}-5}{6\sqrt{37}}\left(\frac{7-\sqrt{37}}{2}\right)^I \text{ for } \mathcal{T}_2\\
\end{cases}
\end{equation}
and of $\mat{B}^{(I)}$ is
\begin{equation}\label{S_B}
S_B(I)=1+\begin{cases}
2\widetilde{S}(I)   \text{ for }  \mathcal{T}_1,\\
\widetilde{S}(I) +\hat{S}(I) \text{ for } \mathcal{T}_2.\\
\end{cases}
\end{equation}

Finally, it is a simple matter to check that for $\mathcal{T}_1$ and $\mathcal{T}_2$ the total number of eigenvalues, $S_B(I) + \widetilde{S}(I) + \sum_{n=1}^ID(n)S(n)$ and $S_B(I) + \widetilde{S}(I) + \sum_{n=1}^I[D(n)S(n)+\hat{D}(n)\hat{S}(n)]$, respectively, is exactly equal to the number of beads at iteration $I$, $N(I)=8^I+1$.  This shows that the constructed sets of eigenmodes are complete.

\begin{figure}[t]
\centering
\includegraphics[width=1.1\linewidth,trim=0 0 0 0]{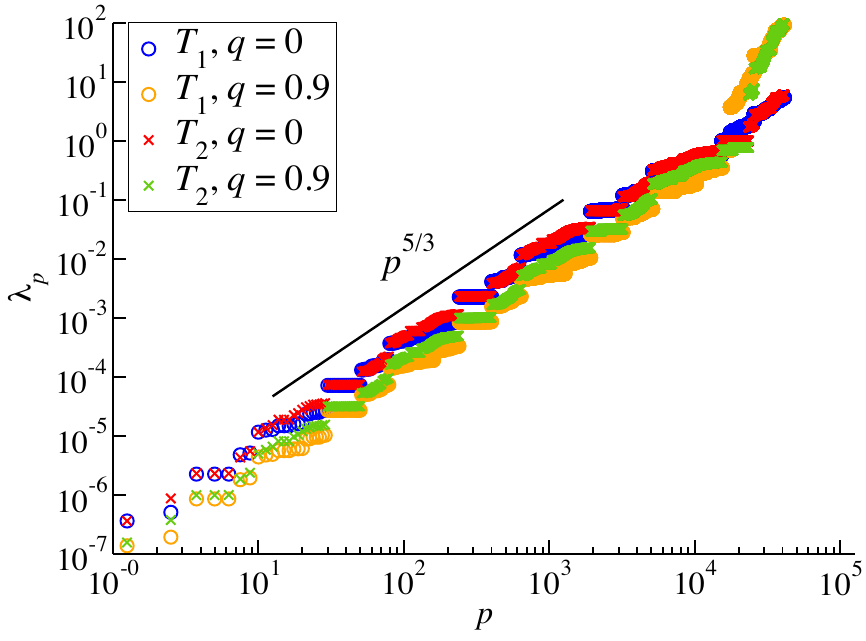}
\caption{Double-logarithmic representation of the eigenvalues of the dynamical matrix of $\mathcal{T}_1$ and $\mathcal{T}_2$ at iteration $I=5$. The parameter $q=0$ is related to fully flexible trees and $q=0.9$ to the semiflexible ones. All spectra show for lower eigenvalues a scaling $\lambda_p\sim p^{5/3}$.}\label{eig}
\end{figure}

In Fig.~\ref{eig} we exemplify the spectra for $\mathcal{T}_1$ and $\mathcal{T}_2$ having stiffness parameter $q=0$ (fully-flexible case) and $q=0.9$ (semiflexible case). As it is typical for semiflexible trees \cite{dolgushev10b,mielke16,grimm16,fuerstenberg12,fuerstenberg13}, switching on the stiffness leads to an increase of higher eigenvalues (due to the restricted local vibrations) and a decrease of the lower ones (due to the growth of the trees' size). Here, the lower eigenvalues scale with the mode number $p$ as $\lambda_p\sim p^{5/3}$, notwithstanding their non-smooth behavior reflecting the degeneracy of eigenvalues. The exponent $5/3$ is directly related to the spectral dimension $d_s=6/5$, $2/d_s=5/3$, that determines the scaling of density of states, $h(\lambda)\sim\lambda^{d_s/2-1}$ \cite{alexander82,dolgushev17a}. Thus, we observe that the local bending rigidity does not affect the spectral dimension.

\begin{table}
\begin{tabular}{l|llll|}
\multicolumn{5}{c}{$q=0$} \\
\cline{2-5} 
$I$ & $\lambda_1(\mathcal{T}_1)$ & $\lambda_2$($\mathcal{T}_1$) & $\lambda_1(\mathcal{T}_2)$ & $\lambda_2$($\mathcal{T}_2$)  \\
\hline
$1$ & $0.2679$ & $0.3446$ & $0.2679$ & $0.5505$   \\
$2$ & $0.01121$ & $0.01549$ & $0.01126$ & $0.02641$  \\
$3$ & $0.0003651$ & $0.0005107$ & $0.0003663$ & $0.0008751$ \\
$4$ & $0.00001147$ & $0.00001607$ & $0.00001150$ & $0.00002754$ \\
$5$ & $3.586\cdot10^{-7}$ & $5.026\cdot10^{-7}$ & $3.597\cdot10^{-7}$ & $8.614\cdot10^{-7}$ \\
\hline
\hline
\end{tabular}
\\
\begin{tabular}{l|llll|}
\multicolumn{5}{c}{$q=0.3$} \\
\cline{2-5}
$I$ & $\lambda_1(\mathcal{T}_1)$ & $\lambda_2$($\mathcal{T}_1$) & $\lambda_1(\mathcal{T}_2)$ & $\lambda_2$($\mathcal{T}_2$)  \\
\hline
$1$ & $0.2082$ & $0.2505$ & $0.2163$ & $0.4648$  \\
$2$ & $0.008302$ & $0.01121$ & $0.008669$ & $0.02039$  \\
$3$ & $0.0002698$ & $0.0003753$ & $0.0002806$ & $0.0006703$ \\
$4$ & $8.477\cdot10^{-6}$ & $0.00001186$ & $8.805\cdot10^{-6}$ & $0.00002108$  \\
$5$ & $2.651\cdot10^{-7}$ & $3.713\cdot10^{-7}$ & $2.753\cdot10^{-7}$ & $6.592\cdot10^{-7}$ \\
\hline
\hline
\end{tabular}
\\
\begin{tabular}{l|llll|}
\multicolumn{5}{c}{$q=0.6$} \\
\cline{2-5}
$I$ & $\lambda_1(\mathcal{T}_1)$ & $\lambda_2$($\mathcal{T}_1$) & $\lambda_1(\mathcal{T}_2)$ & $\lambda_2$($\mathcal{T}_2$) \\
\hline
$1$ & $0.1638$ & $0.1871$ & $0.1765$ & $0.3989$ \\
$2$ & $0.006071$ & $0.007997$ & $0.006629$ & $0.01567$ \\
$3$ & $0.0001966$ & $0.0002716$ & $0.0002131$ & $0.0005091$  \\
$4$ & $6.175\cdot10^{-6}$ & $8.625\cdot10^{-6}$ & $6.680\cdot10^{-6}$ & $0.00001599$ \\
$5$ & $1.931\cdot10^{-7}$ & $2.704\cdot10^{-7}$ & $2.088\cdot10^{-7}$ & $4.999\cdot10^{-7}$ \\
\hline
\hline
\end{tabular}

\begin{tabular}{l|llll|}
 \multicolumn{5}{c}{$q=0.9$} \\
\cline{2-5}
$I$ & $\lambda_1(\mathcal{T}_1)$ & $\lambda_2$($\mathcal{T}_1$) & $\lambda_1(\mathcal{T}_2)$ & $\lambda_2$($\mathcal{T}_2$)  \\
\hline
$1$ & $0.1305$ & $0.1427$ & $0.1453$ & $0.3478$ \\
$2$ & $0.004319$ & $0.005527$ & $0.004990$ & $0.01190$ \\
$3$ & $0.0001385$ & $0.0001897$ & $0.0001585$ & $0.0003789$ \\
$4$ & $4.350\cdot10^{-6}$ & $6.062\cdot10^{-6}$ & $4.963\cdot10^{-6}$ & $0.00001188$ \\
$5$ & $1.360\cdot10^{-7}$ & $1.904\cdot10^{-7}$ & $1.551\cdot10^{-7}$ & $3.713\cdot10^{-7}$ \\
\hline
\hline
\end{tabular}
\caption{First two minimal nonvanishing eigenvalues of the dynamical matrix of fractal trees $\mathcal{T}_1$ and $\mathcal{T}_2$ at various iterations $I$ and for different values of the stiffness parameter $q$.}\label{table}
\end{table}

For many quantities related to global physics the lowest eigenvalues play a major role. Looking at Fig.~\ref{eig} one can observe that the lowest non-vanishing eigenvalue $\lambda_1$ is (almost) equal for $\mathcal{T}_1$ and $\mathcal{T}_2$ in case of $q=0$ and it is slightly higher for $\mathcal{T}_2$ for $q=0.9$, see also Table~\ref{table}.  This eigenvalue comes from the matrix $\tmat{A}^{(I)}$ and related to the eigenmode in which the largest branches move as whole, such as depicted in case (d) of Figs.~\ref{modes_a}-\ref{modes_p} for $I=1$. Going to the second smallest eigenvalue $\lambda_2$, one observes large deviations between the structures, see Table~\ref{table} and Fig.~\ref{eig}. Especially in the semiflexible case ($q=0.9$) the difference is almost given by factor two. Eigenvalue $\lambda_2$ follows from matrix $\mat{B}^{(I)}$ and related to the mode such as displayed in Figs.~\ref{modes_a}-\ref{modes_p}(h). This mode involves motion of side-chains as whole that are longer in case of tree $\mathcal{T}_1$ leading hence for this tree to a smaller $\lambda_2$.

\section{Static and dynamical properties of the trees}\label{properties}


Based only on  the eigenvalues $\{\lambda_p\}$ of the dynamical matrix $\mathbf{A}$ (and not on its eigenvectors), many static and dynamical properties of Gaussian polymers can be readily calculated. First, the size of a polymeric structure is typically characterized by the mean-square gyration radius $\langle R_g^2\rangle\equiv\frac{1}{2N^2}\sum_{i,j}^N \langle(\mathbf{r}_i-\mathbf{r}_j)^2\rangle$, that can be straightforwardly calculated from $\{\lambda_p\}$ \cite{forsman76,eichinger80,dolgushev10b,jurjiu14}:
\begin{equation}\label{R_g}
 \langle R_g^2\rangle=\frac{b^2}{N}\sum_{p=1}^{N-1} \frac{1}{\lambda_p}
\end{equation}
Here and in the following expressions the sum runs over all eigenvalues, except $\lambda_0=0$ related to the motion of the macromolecule as whole.
We note the direct relation of the mean-square gyration radius of fully-flexible Gaussian trees ($q=0$) to Wiener index, $\mathrm{Wi}=N^2\langle R_g^2\rangle/b^2$, see, e.g., Ref. \cite{nitta94}. 

\begin{figure}[t]
\centering
\includegraphics[width=1\linewidth,trim=0 0 0 0]{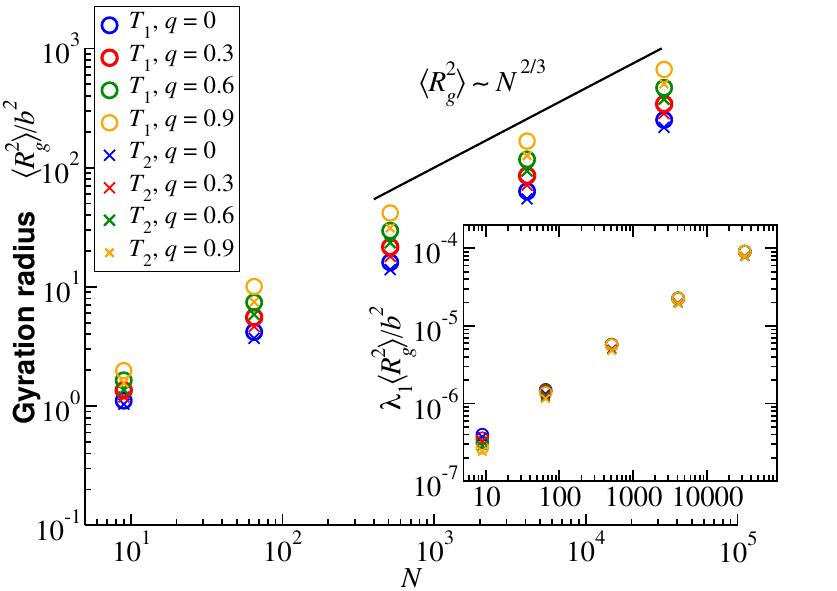}
\caption{Double-logarithmic representation of the mean-square gyration radii $\langle R_g^2\rangle$ of trees $\mathcal{T}_1$ and $\mathcal{T}_2$ as function of total number of beads $N(I)=8^I+1$ for different values of the stiffness parameter $q$. Inset shows rescaled radii with the minimal nonvanishing eigenvalue $\lambda_1$.}\label{Rg}
\end{figure}

In Fig.~\ref{Rg} we plot the mean-square gyration radius $\langle R_g^2\rangle$ of $\mathcal{T}_1$ and $\mathcal{T}_2$ as function of number of beads $N$. As can be inferred from the figure (solid line), both trees are compact for $I\geq3$, i.e. $\sqrt{\langle R_g^2\rangle}\sim N^{1/3}$ \footnote{The exponent $1/3$ is related to the spectral dimension $d_s=6/5$ by $(2-d_s)/2d_s=1/3$, see, e.g., Ref.~\cite{jurjiu14}}.  As one expects, structures with higher stiffness parameter $q$ have higher gyration radius. Given that the lower eigenvalues play a major role for $\langle R_g^2\rangle$, see Eq.~\eqref{R_g}, their dependency on $q$ determines the behavior of $\langle R_g^2\rangle$, see Figs.~\ref{eig} and \ref{Rg}. For a given tuple $(I,q)$ the gyration radius of $\mathcal{T}_1$ is higher than for $\mathcal{T}_2$. Such a behavior is quite expectable from the structure of $\mathcal{T}_1$ that has more beads with a longer topological distance from the core than those of $\mathcal{T}_2$.  This fact corresponds also to lower eigenvalues for the trees, see Fig.~\ref{eig} and Table~\ref{table}. The significant growth of the gyration radii with increasing stiffness is due to the ground states. In the inset to Fig.~\ref{Rg} we show the $\langle R_g^2\rangle$ multiplied by $\lambda_1$ that leads to a collapse of the data for each tree. The points for $\mathcal{T}_2$ remain still under those of $\mathcal{T}_1$ due to the large difference in the second minimal nonvanishing eigenvalue $\lambda_2$, see Table~\ref{table}.

\begin{figure}[t]
\centering\includegraphics[width=1\linewidth]{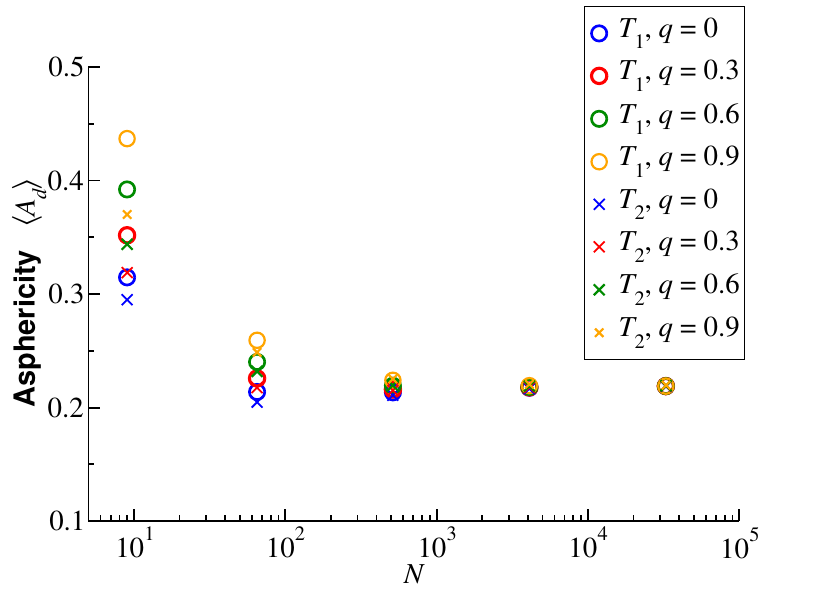}
\centering\includegraphics[width=1\linewidth]{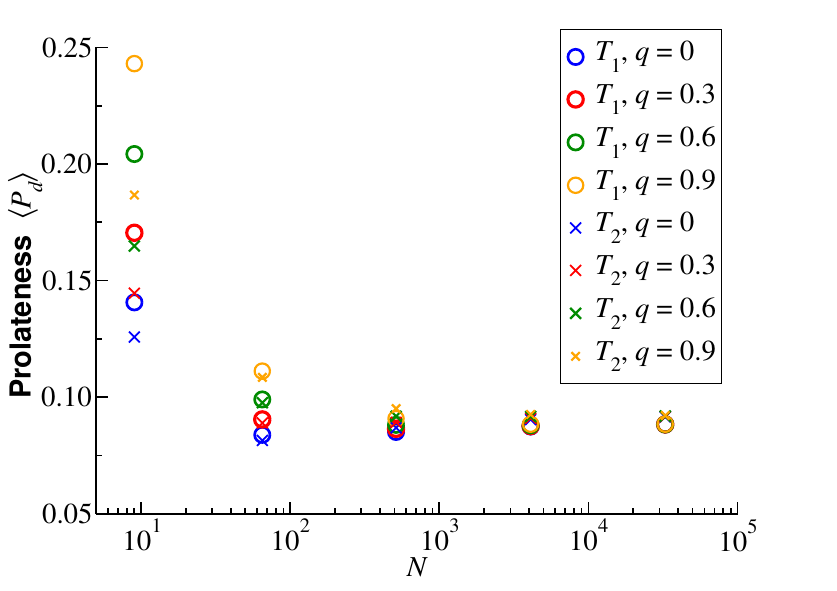}
\caption{Half-logarithmic representation of the asphericity $\langle A_d\rangle$ and prolateness $\langle P_d\rangle$ of trees $\mathcal{T}_1$ and $\mathcal{T}_2$ as function of total number of beads $N(I)=8^I+1$ for different values of the stiffness parameter $q$. As expected for large self-similar objects (here for $N>500$), both characteristics saturate to a plateau.}\label{Shapes}
\end{figure}

The gyration radius does not provide information about deviations from the spherical shape. For this one has to consider the eigenvalues $(\sigma_1,\sigma_2,\sigma_3)$ of the  gyration tensor, such that $\sigma_1>\sigma_2>\sigma_3$ and $\sigma_1+\sigma_2+\sigma_3=R_g^2$ hold. Based on $\{\sigma_i\}$ one commonly calculates \footnote{We note that in simulations one usually calculates preaveraged asphericity and prolateness, i.e., based on the average $\{\protect\langle\sigma_i\protect\rangle\}$ and $\protect\langle R_g^2\protect\rangle$.}  the average asphericity $\langle A_d\rangle$ and prolateness $\langle P_d\rangle$, which in $d=3$ dimension are given by \cite{rudnick86,rudnick87,wei95,wei97a,wei97b,zifferer99,vonferber09,vonferber15,kalyuzhnyi16}
\begin{equation}\label{aspher_def_av}
\langle A_d\rangle=\left\langle\frac{\sum_{i<j}^3(\sigma_i-\sigma_j)^2}{2R_g^4}\right\rangle
\end{equation}
and
\begin{equation}\label{prolate_def_av}
\langle P_d\rangle=\left\langle\frac{\prod_{i=1}^3(3\sigma_i-R_g^2)}{2R_g^6}\right\rangle.
\end{equation}
The limiting values for asphericity $\langle A_d\rangle$ are $0$ for spherical shape and $1$ for rodlike shape. The prolateness $\langle P_d\rangle$ takes negative values from $(-1/8,0)$ for oblate shapes and positive values from $(0,1)$ for prolate shapes. As for asphericity, if prolateness is zero, the shape of the structure is spherical \cite{rudnick86,rudnick87,wei95,wei97a,wei97b,zifferer99,vonferber09,vonferber15,kalyuzhnyi16}. In the dimension $d=3$ the average asphericity $\langle A_d\rangle$ and the average prolateness $\langle P_d\rangle$ read \cite{wei95,wei97a,wei97b,vonferber09,vonferber15}
\begin{equation}\label{aspher}
\langle A_d\rangle=\frac{15}{2}\int_0^{\infty}\mathrm{d}y\sum_{k=1}^{N-1} \frac{y^3}{(\lambda_k+y^2)^2}\left[\prod_{j=1}^{N-1}\frac{\lambda_j}{\lambda_j+y^2}\right]^{\frac{3}{2}}
\end{equation}
and
\begin{equation}\label{prol}
\langle P_d\rangle=\frac{105}{8}\int_0^{\infty}\mathrm{d}y\sum_{k=1}^{N-1} \frac{y^5}{(\lambda_k+y^2)^3}\left[\prod_{j=1}^{N-1}\frac{\lambda_j}{\lambda_j+y^2}\right]^{\frac{3}{2}},
\end{equation}
respectively.

In Fig.~\ref{Shapes} we plot average asphericity $\langle A_d\rangle$ and prolateness $\langle P_d\rangle$ for trees $\mathcal{T}_1$ and $\mathcal{T}_2$ of  different size $N(I)$ and stiffness $q$. First, one can see that both trees have an aspheric shape that for high iterations $I$ saturates to an universal value $\langle A_d\rangle\simeq0.22$ for all considered values of $q$. Thus, the trees are less aspherical than ideal linear chains \cite{rudnick87,kalyuzhnyi16} or combs \cite{wei97b,vonferber15} and more aspherical than ideal stars with $f>4$ arms \cite{wei97b}. Furthermore, both trees $\mathcal{T}_1$ and $\mathcal{T}_2$ are prolate, given that $\langle P_d\rangle>0$. For larger iterations the data collapse for all considered values of $q$ on $\langle P_d\rangle\simeq0.088$ for $\mathcal{T}_1$ and $\langle P_d\rangle\simeq0.092$ for $\mathcal{T}_2$. The latter prolateness value of $0.092$ for $\mathcal{T}_2$ is close to that of the $4$-arm-star \cite{wei97b}. Tree $\mathcal{T}_1$ is less prolate (that is also evident from the topology of the tree, Fig.~\ref{trees}), the corresponding value $0.088$ lies between that of the $4$-arm and $5$-arm-stars \cite{wei97b}.

\begin{figure}[t]
\centering
\includegraphics[width=1\linewidth,trim=0 0 0 0]{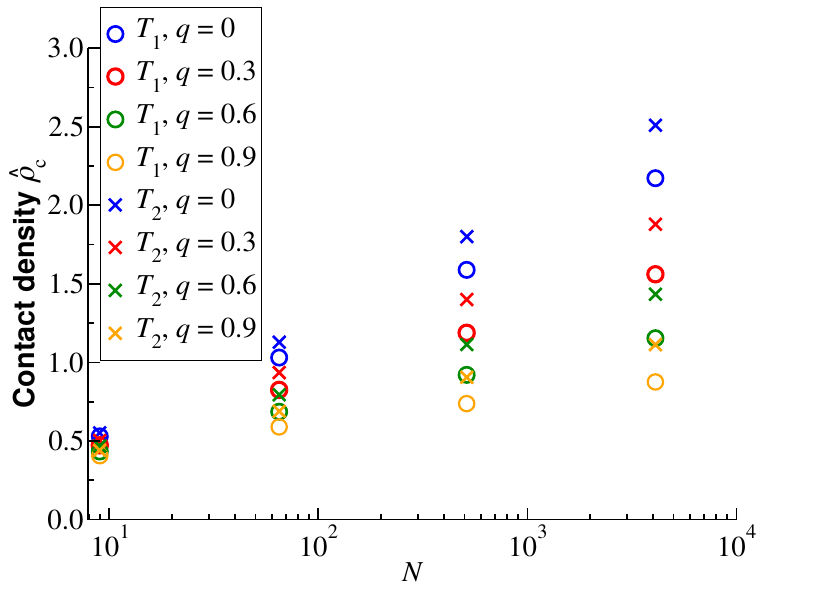}
\caption{Half-logarithmic representation of the number of self-contacts per monomer for trees $\mathcal{T}_1$ and $\mathcal{T}_2$ as function of total number of beads $N(I)=8^I+1$ for different values of the stiffness parameter $q$.}\label{Contacts}
\end{figure}

While the mean-square gyration radius and the shape parameters can be calculated based on the eigenvalues only, for many quantities more information about the structures is needed. Here we consider the equilibrium density of contacts and the form factors of the trees. Both characteristics can be calculated based on the matrix of equilibrium mean-square distances $\mathbf{L}=(L_{ij})$, where $L_{ij}$ gives the mean-square distance between monomers $i$ and $j$ (in the units of $b^2$). The matrix $\mathbf{L}$ is directly related to the (symmetric) dynamical matrix $\mathbf{A}$ by \cite{klein93,fouss07,zhang10}
\begin{equation}
L_{ij}=A^{\dagger}_{ii}+A^{\dagger}_{jj}-2A^{\dagger}_{ij},
\end{equation}
where $\{A^{\dagger}_{ij}\}$ are the elements of the Moore-Penrose pseudo-inverse matrix $\mathbf{A}^{\dagger}$ of $\mathbf{A}$. Given that the singularity of the matrix $\mathbf{A}$ comes from the translational mode $\mathbf{v}_0=(1,1,\dots,1)/\sqrt{N}$ [such as depicted in Figs.~\ref{modes_a}-\ref{modes_p}(i)] that leads to the eigenvalue $\lambda_{0}=0$, the pseudo-inverse of $\mathbf{A}$ can be readily computed, 
\begin{equation}
\mathbf{A}^{\dagger}=(\mathbf{A}-\mathbf{v}_0\otimes\mathbf{v}_0)^{-1}+\mathbf{v}_0\otimes\mathbf{v}_0.
\end{equation}

Now, the probability $p_{ij}$ that two monomers (say, $i$ and $j$) are in contact is given by $p_{ij}=(2\pi L_{ij}/3)^{-3/2}$ \cite{doi88}. With this, the contact density (i.e., number of contacts per monomer) reads
\begin{equation}
\hat{\rho}_c=\frac{1}{N}\sum_{i<j}\left(\frac{3}{2\pi L_{ij}}\right)^{3/2}.
\end{equation}
In Fig.~\ref{Contacts} we show the contact density for different values of stiffness parameter $q$. Introducing stiffness leads to a tremendous reduction of the number of contacts. Moreover, this effect is more striking for larger trees. For fully-flexible ($q=0$) tree $\mathcal{T}_1$ at iteration $I=4$ the number of contacts per bead is higher than two, whereas introducing semiflexibility to this tree leads, e.g. for $q=0.9$, to less than one contact per bead. Generally, tree $\mathcal{T}_1$ has lower contact density than $\mathcal{T}_2$ of the same size $N$ and stiffness $q$. This observation corresponds to the higher gyration radius of $\mathcal{T}_1$ in comparison to $\mathcal{T}_2$, see Fig.~\ref{Rg}.

\begin{figure}[t]
\centering
\includegraphics[width=1.1\linewidth,trim=0 0 0 0]{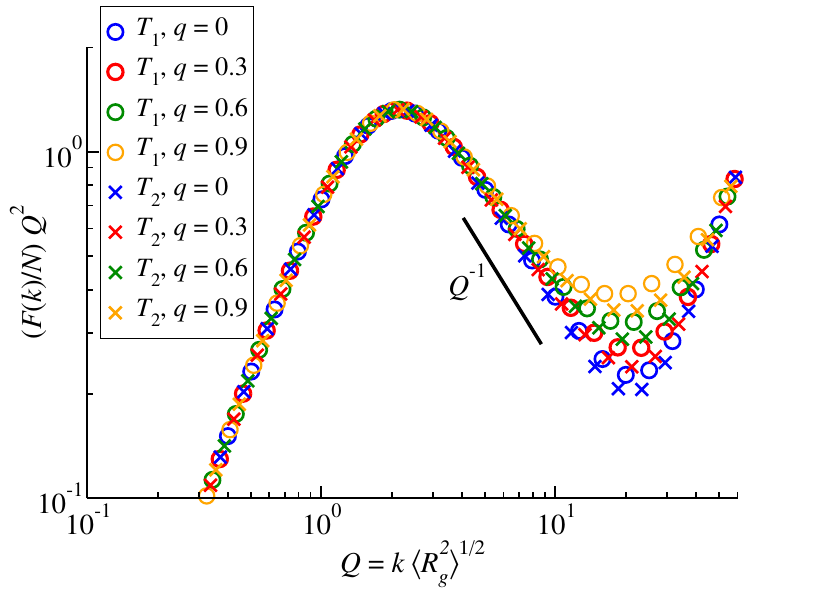}
\caption{Kratky representation $k^2F(k)$ of the form factor of trees $\mathcal{T}_1$ and $\mathcal{T}_2$ at iteration $I=4$ for different values of the stiffness parameter $q$. In the intermediate region of $Q=k\sqrt{\langle R_g^2\rangle}$, the data saturate on the scaling $Q^{-1}$.
}\label{Form_factor}
\end{figure}

The internal organization of macromolecules is studied in scattering experiments by looking at the coherent intramolecular form factor $F(k)={\frac{1}{N}\sum_{i,j}^N\langle\exp[\mathrm{i}\mathbf{k}\cdot(\mathbf{r}_i-\mathbf{r}_j)]\rangle}$. For Gaussian distributed $\{\mathbf{r}_i\}$ the form factor $F(k)$ can be formulated in terms of the distance matrix $\mathbf{L}$  \cite{doi88},
\begin{equation}
F(k)=\frac{1}{N}\sum_{i,j}^N\exp\left[-\frac{k^2b^2}{6}L_{ij}\right].
\end{equation}
In Fig.~\ref{Form_factor} we plot the form factor of trees $\mathcal{T}_1$ and $\mathcal{T}_2$ at iteration $I=4$ for different values of the stiffness parameter $q$ using Kratky representation. Moreover, we rescale the wave vector by taking $Q=k\sqrt{\langle R_g^2\rangle}$. In this representation all data for $Q\lesssim4$ collapse. For higher $Q$ the data for stiffer structures lie above those of the flexible ones, reflecting more swollen local organization of the trees. In the intermediate region of $1<Q<10$ the data approach scaling, $F(k)\sim k^{-3}$. The differences at rather large $Q\simeq10$ reflect their local character, hence for higher iterations $I$ they are expected to be less relevant.

\begin{figure}[t]
\centering
\includegraphics[width=1.1\linewidth,trim=0 0 0 0]{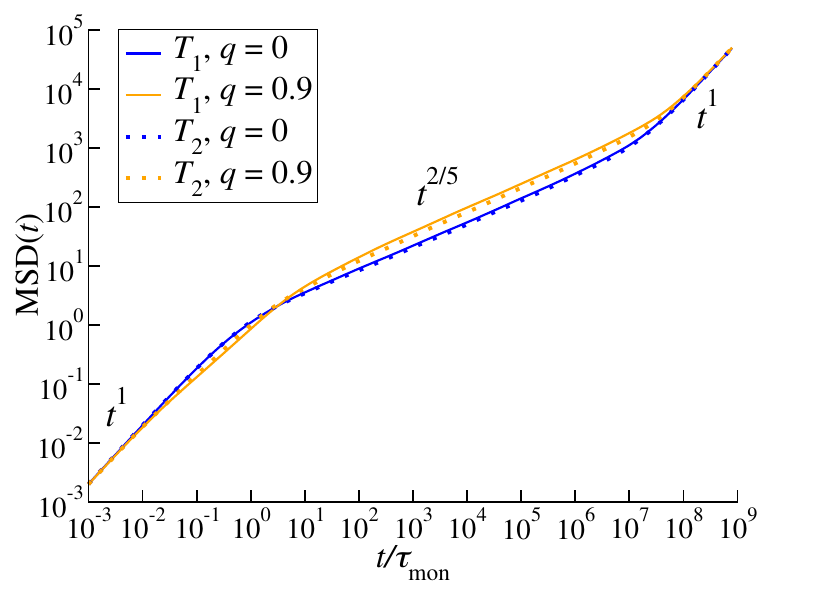}
\caption{Double-logarithmic representation of the monomeric MSD for trees $\mathcal{T}_1$ and $\mathcal{T}_2$ at iteration $I=5$ having different values of stiffness parameter $q$.}\label{MSD}
\end{figure}

We close the discussion of static properties of the trees and proceed to the dynamics of the structures. First, we consider the mean-square displacement (MSD) of monomers averaged over the whole structure, that follows from Eq.~\eqref{langevin} and is given by \cite{doi88,gurtovenko05}
\begin{equation}\label{MSD_eq}
\overline{\langle(\mathbf{r}(t)-\mathbf{r}(0))^2\rangle}=\frac{2b^2}{N}\left( \frac{t}{\tau_{\mathrm{mon}}}+\sum_{p=1}^{N-1}\frac{1-e^{-t\lambda_p/\tau_{\mathrm{mon}}}}{\lambda_p}\right),
\end{equation} 
where $\langle\cdots\rangle$ and $\overline{\cdots}$ denote conformational and structural averages, respectively, and $\tau_{\mathrm{mon}}=\zeta b^2/3k_BT$ is the monomeric relaxation time. The results for MSD of the trees at iteration $I=5$ are presented in Fig.~\ref{MSD}. Apart from evident scaling $t^1$ for $t\ll\tau_{\mathrm{mon}}$ and $t\gg\tau_{\mathrm{mon}} N^{5/3}$, there is subdiffusion $t^{2/5}$ at intermediate times. The exponent $2/5$ is closely related to the spectral dimension $d_s=6/5$ \cite{dolgushev17a}: The relation $2/5=1-d_s/2$ follows straightforwardly from Eq.~\eqref{MSD_eq} if one replaces there the sum through an integral, $\sum\dots\rightarrow\int\mathrm{d}\lambda h(\lambda)\dots$, where $h(\lambda)\sim\lambda^{d_s/2-1}$ is the density of states. The subdiffusive exponent is robust under introduction of stiffness, the MSD of beads belonging to stiffer structures is slightly higher at intermediate times.

\begin{figure}[t]
\centering
\includegraphics[width=1.1\linewidth,trim=0 0 0 0]{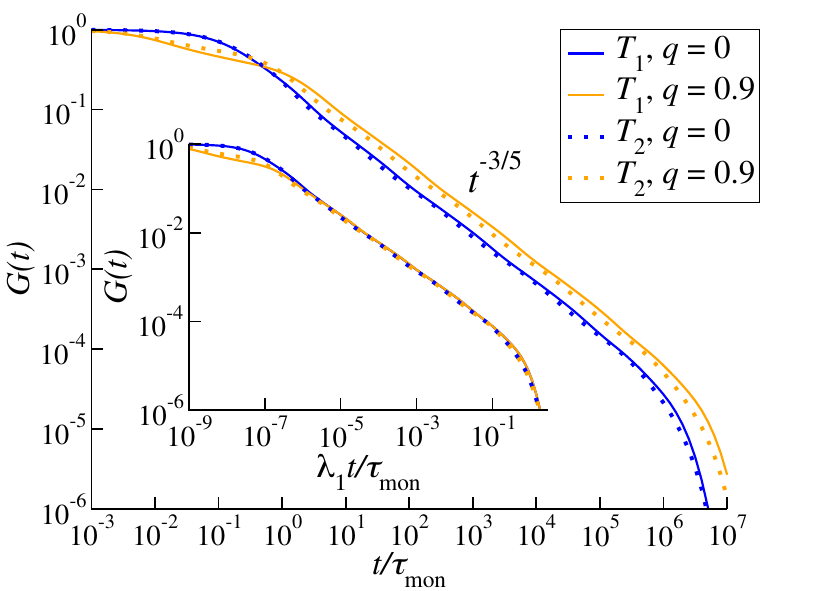}
\includegraphics[width=1.1\linewidth,trim=0 0 0 0]{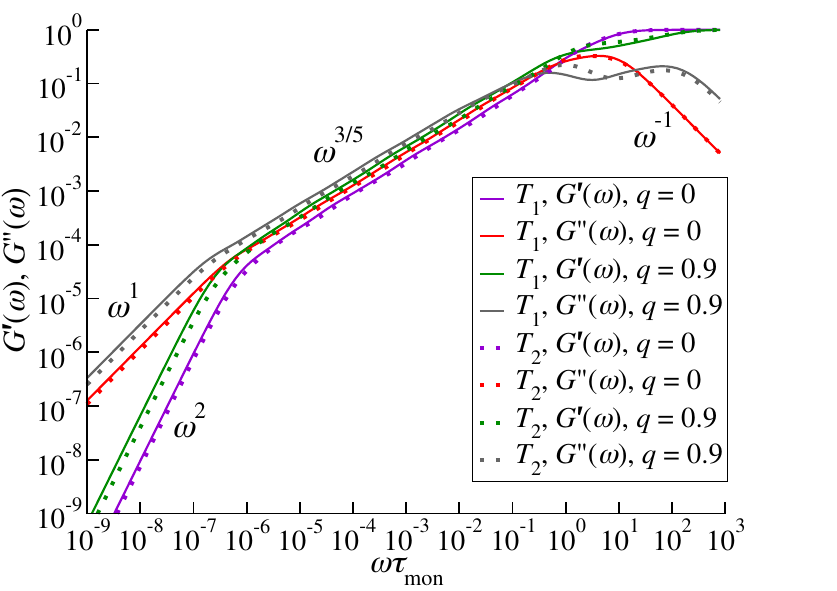}
\caption{(top) Double-logarithmic representation of the shear-stress relaxation modulus $G(t)$ for trees $\mathcal{T}_1$ and $\mathcal{T}_2$ at iteration $I=5$ having different values of stiffness parameter $q$. Inset shows $G(t)$ with rescaled by factor $\lambda_1$ time. (bottom) Double-logarithmic representation of the storage $G'(\omega)$ and loss $G''(\omega)$ moduli corresponding to $G(t)$ of the main top plot. For all curves $ck_BT=1$ is taken.}\label{Gt}
\end{figure}

In the mechanical relaxation experiments one measures responses to external strain fields. The typical response function is the shear relaxation modulus that follows for Gaussian macromolecules the relation \cite{eichinger80,doi88,gurtovenko05}
\begin{equation}\label{Gt_eq}
 G(t)=\frac{ck_BT}{(N-1)}\sum_{p=1}^{N-1} \exp\left[-\frac{2\lambda_p t}{\tau_{\mathrm{mon}}}\right],
\end{equation} 
where $c$ is the number density of the segments. The development of $G(t)$ with time is exemplified for trees $\mathcal{T}_1$ and $\mathcal{T}_2$ on Fig.~\ref{Gt}. Also there we plot experimentally relevant frequency representatives of $G(t)$, the storage $G'(\omega)$ and loss $G''(\omega)$ moduli \cite{doi88}, 
\begin{equation}\label{Gstorage}
 G'(\omega)=\frac{ck_BT}{(N-1)}\sum_{p=1}^{N-1}\frac{(\omega\tau_{\mathrm{mon}}/2\lambda_p)^2}{1+(\omega\tau_{\mathrm{mon}}/2\lambda_p)^2}
\end{equation} 
and
\begin{equation}\label{Gloss}
G''(\omega)=\frac{ck_BT}{(N-1)}\sum_{p=1}^{N-1}\frac{\omega\tau_{\mathrm{mon}}/2\lambda_p}{1+(\omega\tau_{\mathrm{mon}}/2\lambda_p)^2}.
\end{equation} 
The initial value of the shear-stress relaxation modulus, $G(t=0^+)=ck_BT$, is given by the affine shear elasticity of a system of ideal springs \cite{wittmer16}.
At the intermediate times the $G(t)$ decays algebraically (here with the exponent $-3/5=-d_s/2$) that readily follow from the behavior of the density of states $h(\lambda)$ \cite{dolgushev17a}. At long times due to the finite size of structures one gets an exponential cut-off related to the minimal eigenvalue $\lambda_1$, see Table~\ref{table}. Exceptionally at initial times, the $G(t)$ for semiflexible ($q=0.9$) trees decays faster than that of the flexible trees $q=0$. (One finds corresponding deviations for $G'(\omega)$ or $G''(\omega)$ at high frequencies.) This behavior shows fast local vibrations in semiflexible trees due to the locally restricted bonds, that are also manifested in the eigenvalues for large mode number $p$ in Fig.~\ref{eig}. Correspondingly to the behavior of $G(t)$, at very low frequencies, $\omega\tau_{\mathrm{mon}}\ll N^{-5/3}$, one finds $G'(\omega)\sim\omega^2$ and $G''(\omega)\sim\omega$; at very high frequencies one has $G'(\omega)\rightarrow ck_BT$ and $G''(\omega)\sim\omega^{-1}$ \cite{doi88}. Moreover, as one expects for self-similar fractal objects of spectral dimension $d_s$, we find in the intermediate frequency regime that
\begin{equation}
G'(\omega) \approx G''(\omega) \sim \omega^{d_s/2}.
\end{equation}

\section{Summary and Conclusions}\label{conclusions}

In summary, in this work we have studied marginally compact trees that are created by means of two fractal generators. We focused on the role of local stiffness for the typical static and dynamical characteristics of the trees. We have shown that introduction of stiffness leads to an increase of size $R$ of the structures. Nevertheless the structures remain compact, by showing a $R\sim N^{1/3}$ scaling. Moreover, the static form factor approaches for large structures an intermediate $F(k)\sim k^{-3}$ behavior. The ensuing exponent can be assigned, from one side, to the fractal dimension $d_f=3$ and, from another side, to a fractal surface with dimension $d_A=3$. (We remind that the objects with a smooth surface, e.g., a ball, have $d_A=2$.) Furthermore, the shape of the trees is not spherical and the corresponding asphericity and prolateness  parameters for large enough structures are independent of the stiffness and the tree structure. At the same time the semiflexibility influences tremendously the density of self-contacts that gets drastically reduced with growing stiffness. In the dynamics, the scaling of the relaxation times, $\tau_p\sim(N/p)^{5/3}$, is reflected in the monomeric mean-square displacement or in the shear-stress relaxation modulus by showing at intermediate times the behavior $t^{2/5}$ or $t^{-3/5}$, respectively.

Coming back to recent paper \cite{dolgushev17a} by some of us, where we have shown that the linear spacers reduce the number of contacts, here we have suggested another recipe for suppression of the self-contact density by introducing local stiffness. We note that so far these findings were demonstrated for ideal trees. In this respect it will be interesting to look on the excluded volume and finite extensibility effects in the future.

\section*{Acknowledgments}
The authors thank A. Blumen and J.-U. Sommer for fruitful discussions.
M.D. acknowledges DFG through GRK 1642/1.

\appendix

\section*{Appendix: Structure of eigenmodes and corresponding reduced matrices}

\subsection{Tree $\mathcal{T}_1$}

\subsubsection{Number of distinct amplitudes}

\begin{figure}[t]
\centering
\includegraphics[width=1\linewidth,trim=0 0 0 0]{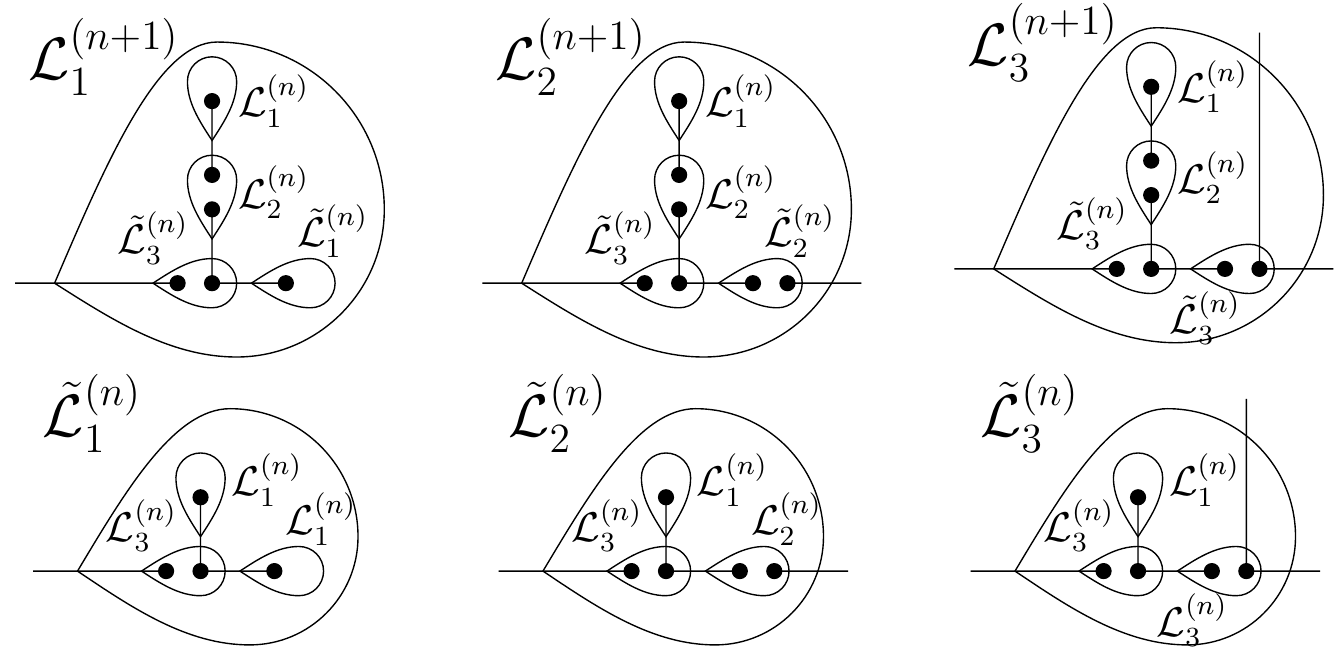}
\caption{Schematic sketch of substructures ("leaves" $\mathcal{L}$) of tree $\mathcal{T}_1$.}\label{leaves_a}
\end{figure}

As has been discussed in the main part of the paper, the symmetry of $\mathcal{T}_1$ allows a construction of eigenmodes in which some beads move with the same amplitude. The number of the distinct non-vanishing amplitudes determines then the size of reduced matrices, that are, e.g., for $I=1$ presented by Eqs.~\eqref{A1}-\eqref{B1}. Now, at higher iterations $I$ one gets a similar pattern of motion as in Fig.~\ref{modes_a}(a)-(b), where two directly connected substructures (called "leaves", see $\mathcal{L}_1^{(n+1)}$ in Fig.~\ref{leaves_a}) move against each other. The modes of Fig.~\ref{modes_a}(c)-(d) bring forth at iteration $I$ the pattern in which two leaves $\mathcal{\tilde {L}}_1^{(I)}$ (see Fig.~\ref{leaves_a}) move against each other. Each such a leave, $\mathcal{L}_1$ or $\mathcal{\tilde {L}}_1$, can be constructed in an iterative way from other leaves $\mathcal{L}_i$ or $\mathcal{\tilde {L}}_i$ (index $i$ indicates that $i-1$ outer leaves are connected to $\mathcal{L}_i$ or to $\mathcal{\tilde {L}}_i$), see  Fig.~\ref{leaves_a}. This construction allows to calculate the number of distinct amplitudes $S(n)$ or $\widetilde{S}(n)$ in the modes involving leaves $\mathcal{L}_1^{(n)}$ or $\mathcal{\tilde {L}}_1^{(n)}$, that give the size of matrices $\mat{A}^{(n)}$ or $\tmat{A}^{(n)}$, respectively.

We start by looking at $\widetilde{S}(n)$. The corresponding leaf $\mathcal{\tilde {L}}_1^{(n)}$ consists of one leaf $\mathcal{L}_3^{(n)}$ and two $\mathcal{L}_1^{(n)}$. There, the beads of one leaf $\mathcal{L}_1^{(n)}$ move with exactly the same amplitude and phase as by their symmetric counterparts in the other leaf $\mathcal{L}_1^{(n)}$ (see Fig.~\ref{modes_a}(c)-(d) for $n=1$). Therefore, the presence of the second leaf $\mathcal{L}_1^{(n)}$ does not increase $\widetilde{S}(n)$. Denoting by $S'(n)$ the number of independent amplitudes coming from $\mathcal{L}_3^{(n)}$, we then get
\begin{align}
	\widetilde{S}(n) = S(n) + S'(n).	\label{tildeS}
\end{align}
In a similar way, by looking at $\mathcal{L}_1^{(n+1)}$ in Fig.~\ref{leaves_a} and using Eq.~\eqref{tildeS}, we obtain the number $S(n)$ of independent amplitudes coming from $\mathcal{L}_1^{(n)}$,
\begin{align}
	S(n) = 3S(n-1) + 4S'(n-1),
	\label{eq:f1n}
\end{align}
where we have used that leaves $\mathcal{L}_2^{(n-1)}$ and $\mathcal{L}_3^{(n-1)}$ bring the same number of independent amplitudes $S'(n)$. Equations~\eqref{tildeS} and \eqref{eq:f1n} involve $S'(n)$, for which the recurrent equation 
\begin{align}
	S'(n) = 3S(n-1) + 5S'(n-1)
	\label{eq:f2n}
\end{align}
holds, as can be found by inspecting $\mathcal{L}_2^{(n+1)}$ or $\mathcal{L}_3^{(n+1)}$ of Fig.~\ref{leaves_a}.

In order to solve the set of recurrent Eqs.~\eqref{tildeS}-\eqref{eq:f2n}, we first subtract \eqref{eq:f1n} from \eqref{eq:f2n}, $S'(n) - S(n) = S'(n-1)$,  from which follows that
\begin{align}
	S'(n)=\sum_{i=1}^n S(i)
	\label{eq:rekf2}
\end{align}
and that
\begin{align}
		S(n) &= 8S(n-1) - 3S(n-2).\label{eqS}
\end{align}
The solution of Eq.~\eqref{eqS} with initial conditions $S(1) = 1$ and $S(2) = 7$ is
\begin{align}
S(n) =\frac{\sqrt{13}-1}{6\sqrt{13}}(4+\sqrt{13})^n+\frac{\sqrt{13}+1}{6\sqrt{13}}(4-\sqrt{13})^n.
\end{align}
Based on this result and upon employment of Eqs.~\eqref{tildeS} and \eqref{eq:rekf2}, the other quantities $\widetilde{S}(n)$ and $S'(n)$ can be readily calculated (the result for $\widetilde{S}(n)$ is given in Eq.~\eqref{tildeSm} of the main text). 

Finally, we discuss the size of matrix $\mat{B}^{(I)}$, that is coming from the modes in which all beads are moving. Here helps the observation that tree $\mathcal{T}_1$ at iteration $I$ consists from two equivalent leaves $\tilde{\mathcal{L}}_1^{(I)}$ that are connected through the core bead to leaf $\mathcal{L}_2^{(I)}$, which is also then connected to a leaf $\mathcal{L}_1^{(I)}$. With this the size of $\mat{B}^{(I)}$, $S_B(I)$, reads:
\begin{align}
	S_B(I) &= \widetilde{S}(I) + S(I) + S'(I) + 1
	\label{eq:fb}
\end{align}
so that, together with Eq.~\eqref{tildeS}, Eq.~\eqref{S_B} for $\mathcal{T}_1$ follows.

\subsubsection{Initial Matrices}

Starting with $I\geq 2$ the next-nearest neighboring (NNN) interactions affect only directly connected leaves. Thus, it is sufficient to initialize iterative construction of reduced matrices based on 
\begin{align}\nonumber
\mat{A}^{(2)}=
\begin{pmatrix}
\mu_3^{(1)}+\rho_2 & \nu_{13}  & 0     & 0      & \rho_3 & 0      & 0 \\
2\nu_{13} 		& \mu_{113}^{(3)}  & 0     & \rho_3 & \nu_{33}  & 0      & \rho_3 \\
0 			& 0 	  & \mu_2^{(1)} & \nu_{13}  & \rho_2 & 0      & 0 \\
0 			& \rho_3 & \nu_{12} & \mu_{13}^{(2)}  & \nu_{23}  & 0      & \rho_3 \\
2\rho_3 	& \nu_{33}  & \rho_2 & \nu_{23}  & \mu_{233}^{(3)}  & \rho_3 & \nu_{33} \\
0 			& 0      & 0     & 0      & \rho_3 & \mu_3^{(1)}  & \nu_{13} \\
0 			& \rho_3 & 0     & \rho_3 & \nu_{33}  & \nu_{13}  & \mu_{133}^{(3)} - \rho_3
\end{pmatrix}
\end{align}
related to an antiphase motion of two neighboring $\mathcal{L}_1^{(2)}$-leaves and on the auxiliary matrix
\begin{align}\nonumber
\mat{H}_1^{(2)}=
\begin{pmatrix}
\mu_{333}^{(3)} & \rho_3 & \nu_{33} & 0 & 0 & \rho_3 & 0 & 0 \\
\rho_3 & \mu_3^{(1)} & \nu_{13} & 0 & 0 &\rho_3 & 0 & 0 \\
\nu_{33} & \nu_{13} & \mu_{13}^{(2)} & 0 & \rho_3 & \nu_{33} & 0 & \rho_3 \\
0 & 0 &0 & \mu_2^{(1)} & \nu_{12} & \rho_2 & 0  & 0 \\
0 & 0 &\rho_3 & \nu_{12} & \mu_{13}^{(2)} & \nu_{23} & 0 & \rho_3 \\
\rho_3 & \rho_3 & \nu_{33} & \rho_2 & \nu_{23} & \mu_{233}^{(3)} &\rho_3 & \nu_{33} \\
0 & 0 & 0 & 0 & 0 & \rho_3 & \mu_3^{(1)} & \nu_{13} \\
0 & 0 & \rho_3 & 0 & \rho_3 & \nu_{33} & \nu_{13} & \mu_{133}^{(3)} - \rho_3
\end{pmatrix}.
\end{align}

\subsubsection{Construction of $\tmat{A}^{(I)}$}

By investigation of Fig.~\ref{leaves_a}, one can see that an $\mathcal{\tilde L}_1$-leaf is formed by three subunits, from which two are symmetrically equivalent. Consequently, they are described by the same matrix. Therefore, the corresponding matrix $\tmat{A}^{(I)}$ has the shape
\begin{align}\label{matAIad}
\tmat{A}^{(I)}=
\begin{pmatrix}
\bm{\upalpha} && \mat{C}_{12} \\
\mat{C}_{21} && \bm{\upbeta}
\end{pmatrix}.
\end{align} 
Here $\bm{\upalpha}$ represents the co-phase movement of two  $\mathcal{L}_1$-leaves, which makes it very similar to the matrix $\mat{A}^{(I)}$: The only difference is in the last diagonal element describing the amplitude numbered by $S(I)$:
\begin{align}
(\bm{\upalpha})_{ij} = 
\begin{cases}
\mu_{133}^{(3)} + \rho_3 &  (i,j) = (S(I),S(I)) \\
(\mat{A}_1^{(I)})_{ij} & \text{ otherwise}
\end{cases}
\end{align}
Furthermore, in Eq.~\eqref{matAIad} $\bm{\upbeta}$ describes the dynamics of the less symmetric $\mathcal{L}_3$-leaf,
\begin{align}
(\bm{\upbeta})_{ij} =
\begin{cases}
	\mu_{133}^{(3)} & (i,j) = (\widetilde{S}'(I-1), \widetilde{S}'(I-1)) \\
	(\mat{H}_1^{(I)})_{ij} & \text{otherwise}
\end{cases}
\end{align}
where $\widetilde{S}'(I-1)=\widetilde{S}(I-1)+S'(I-1)$ and
\begin{align}
\mat{H}_1^{(I)} = 
\begin{pmatrix}
	\bm{\upzeta}	&& 0 &&	\mat{E}_{13} && \mat{E}_{14} \\
	0 && \bm{\delta} && \mat{E}_{23} && 0 \\
	\mat{E}_{13}^{ \mathrm{T}}  && \mat{E}_{23}^{ \mathrm{T}}  && \bm{\upepsilon} && \mat{E}_{34} \\
	\mat{E}_{14}^{\mathrm{T}}  && 0 && \mat{E}_{34}^{ \mathrm{T}} && \bm{\upzeta}
\end{pmatrix}.
\end{align}
Here $\bm{\upzeta}$ stands for the antiphase movement of two $\mathcal{L}_3$-leaves and $\bm{\updelta}$ or $\bm{\upepsilon}$ describe isolated (i.e., that do not have a symmetrically equivalent neighboring partner) leaves $\mathcal{L}_1$ or $\mathcal{L}_2$, respectively.  The structure of these blocks is provided in Eqs.~\eqref{matdelta}, \eqref{matepsilon}, and \eqref{matzeta}, \textit{vide infra}. The interactions between these leaves are described by very sparse matrices $\mat{E}_{ji}=\mat{E}_{ij}^{ \mathrm{T}}$:
\begin{align}\nonumber
(\mat{E}_{13})_{ij} &=
\begin{cases}
	\rho_3 & (i,j) = (\widetilde{S}'(I), S'(I)) \\
	0 & \text{otherwise}
\end{cases}
\end{align}    
\begin{align}\nonumber
(\mat{E}_{14})_{ij} &= 
\begin{cases}
	\nu_{33} &  (i,j) =  (\widetilde{S}'(I),1) \\
	\rho_3 & (i,j) \in  \{(\widetilde{S}'(I),3), (\widetilde{S}'(I)-1,1),\\ 
	& \hspace{1.4cm}(\widetilde{S}'(I)-2,1)\} \\
	0 & \text{otherwise} 
\end{cases} 
\end{align}    
\begin{align}\nonumber
(\mat{E}_{23})_{ij}&=
\begin{cases}
	\nu_{23} & (i,j) = (S(I),1) \\
	\rho_2 & (i,j) = (S(I),3) \\
	\rho_3 & (i,j) \in \{(S(I)-1,1), (S(I)-2,1)\}\\
	0 & \text{otherwise}
\end{cases} 
\end{align}    
\begin{align}\nonumber
(\mat{E}_{34})_{ij}&=
\begin{cases}
	\nu_{33} & (i,j) = (S'(I),1) \\
	\rho_3 & (i,j) \in \{(S'(I),3), (S'(I)-1,1),\\ 
	& \hspace{1.4cm}(S'(I)-2,1)\}\\
	0 & \text{otherwise}
\end{cases}
\end{align}

Finally, the off-diagonal matrices $\mat{C}_{12}$ and $\mat{C}_{21}$ in Eq.~\eqref{matAIad} are
\begin{align}\nonumber
(\mat{C}_{12})_{ij}&=
\begin{cases}
	\nu_{33} & (i,j) = (S(I),1) \\
	\rho_3 & (i,j) \in\{ (S(I)-1,1), (S(I)-2,1),\\ 
	& \hspace{1.4cm}(S(I),3)\} \\
	0 & \text{otherwise}
\end{cases}
\end{align} 
and  $\mat{C}_{21}=2\mat{C}_{12}^{ \mathrm{T}}$.

\subsubsection{Construction of $\mat{A}^{(I+1)}$}

As can be inferred from Fig.~\ref{modes_a}, leaf $\mathcal{L}_1$ consists from leaves $\mathcal{\tilde L}_1$, $\mathcal{L}_1$, $\mathcal{L}_2$ and $\mathcal{\tilde L}_3$ of the previous iteration. With this the matrix describing antiphase motion of two directly connected $\mathcal{L}_1^{(I+1)}$-leaves is given by
\begin{align}\label{matAad}
\mat{A}^{(I+1)} = 
\begin{pmatrix}
 \bm{\upgamma} && 0 && \mat{D}_{13} && \mat{D}_{14} \\
 0 && \bm{\updelta} && \mat{D}_{23} && 0 \\
 \mat{D}_{12}\transp && \mat{D}_{23}\transp && \bm{\upepsilon} && \mat{D}_{34} \\
 \mat{D}_{14}\transp && 0 && \mat{D}_{34}\transp && \bm{\upzeta}
\end{pmatrix}.
\end{align}
Here $\bm{\upgamma}$ describes the movement of an isolated $\mathcal{\tilde L}_1^{(I)}$-leaf. With a small modification concerning its last bead having number $\widetilde{S}(I)$, we can obtain an expression for $\bm{\upgamma}$:
\begin{align}\label{matgamma}
(\bm{\upgamma})_{ij} = 
\begin{cases}
	\mu_{133}^{(3)} & (i,j) =  (\widetilde{S}(I),\widetilde{S}(I)) \\
	(\tmat{A}^{(I)})_{ij} &\text{otherwise}
\end{cases}
\end{align}
The other matrices $\bm{\updelta}$, $\bm{\upepsilon}$ and $\bm{\upzeta}$ standing for the remaining $\mathcal{L}_1$-, $\mathcal{L}_2$- and $\mathcal{\tilde L}_3$-leaves, respectively, can be constructed as follows. Matrix $\bm{\updelta}$ describes the dynamics of an isolated $\mathcal{L}_1$-leave, in a similar fashion as for $\bm{\upgamma}$, $\bm{\updelta}$ follows from $\mat{A}^{(I)}$:
\begin{align}\label{matdelta}
(\bm{\updelta})_{ij} = 
\begin{cases}
	\mu_{123}^{(3)} & (i,j) =  (S(I),S(I)) \\
	(\mat{A}^{(I)})_{ij} &\text{otherwise}
\end{cases}
\end{align}
Matrix $\bm{\upepsilon}$ reflects the dynamics of an isolated $\mathcal{L}_2$-leave, which is less symmetric than $\mathcal{\tilde L}_1$ or $\mathcal{L}_1$. Its similarity to an $\mathcal{L}_3$-leaf makes it possible to reuse the helper matrix $\mat{H}_1^{(I-1)}$:
\begin{align}\label{matepsilon}
(\bm{\upepsilon})_{ij}=
\begin{cases}
	\mu_{33}^{(2)} & (i,j) = (1,1) \\
	\mu_{123}^{(3)} & (i,j) = (3,3) \\
	\nu_{23} & (i,j) \in\{ (1,3),(3,1)\} \\
	\mu_{133}^{(3)} & (i,j) = (S'(I), S'(I)) \\
	(\mat{H}_1^{(I-1)})_{ij} & \text{otherwise}
\end{cases}
\end{align}
Finally, the $\mathcal{\tilde L}_3$-leaf represented by $\bm{\upzeta}$ in Eq.~\eqref{matAad} has a high similarity to the previously discussed $\mathcal{\tilde L}_1$-leaf. We introduce another helper matrix 
\begin{align}
\mat{H}_2^{(I)} =
\begin{pmatrix}
\bm{\upbeta} && \mat{F}_{12} && \mat{F}_{13}\\
\mat{F}_{12}^{ \mathrm{T}} && \bm{\upalpha} && \mat{F}_{23} \\
\mat{F}_{12}^{ \mathrm{T}} && \mat{F}_{23}^{ \mathrm{T}} && \bm{\upbeta}
\end{pmatrix}.
\end{align}
With only one small modification one can now obtain $\bm{\upzeta}$:
\begin{align}\label{matzeta}
(\bm{\upzeta})_{ij} = 
\begin{cases}
	\mu_{133}^{(3)} & (i,j) \in\{ (S'(I-1),S'(I-1)),\\ 
	& (\tilde S(I-1),\tilde S(I-1) \}\\
	(\mat{H}_2^{(I)})_{ij} & \text{otherwise}
\end{cases}
\end{align}
The interaction matrices follow readily, keeping in mind that $\mat{F}_{ij}=\mat{F}_{ji}^{ \mathrm{T}}$ and $\mat{D}_{ij}=\mat{D}_{ji}^{ \mathrm{T}}$,
\begin{align}\nonumber
(\mat{F}_{12})_{ij}&=
\begin{cases}
	\rho_3 & (i,j) = (S'(I),S(I)) \\
	0 & \text{otherwise}
\end{cases} 
\end{align}    
\begin{align}\nonumber
(\mat{F}_{13})_{ij}&=
\begin{cases}
	\nu_{33} & (i,j) = (S'(I),1) \\
	\rho_3 & (i,j) \in \{ (S'(I),3),(S'(I)-1,1),\\ 
	& \hspace{1.4cm}(S'(I)-2,1)\} \\
	0 & \text{otherwise}
\end{cases} 
\end{align}    
\begin{align}\nonumber
(\mat{F}_{23})_{ij}&=
\begin{cases}
	\nu_{33} & (i,j) = (S(I),1) \\
	\rho_3 & (i,j) \in\{ (S(I),3),(S(I)-1,1),\\ 
	& \hspace{1.4cm}(S(I)-2,1)\} \\
	0 & \text{otherwise}
\end{cases}
\end{align}
and
\begin{align}\nonumber
(\mat{D}_{13})_{ij} =
\begin{cases}
	\rho_3 & (i,j) = (\widetilde{S}(I), S'(I)) \\
	0 & \text{otherwise}
\end{cases}
 \end{align}    
\begin{align}\nonumber
(\mat{D}_{14})_{ij} =
\begin{cases}
	\nu_{33} & (i,j) = (\widetilde{S}(I),1) \\
	\rho_3 & (i,j) \in\{ (\widetilde{S}(I)-1, 1),(\widetilde{S}(I)-2, 1),\\ 
	& \hspace{1.4cm}(\widetilde{S}(I), 3)\} \\
	0 & \text{otherwise}
\end{cases}
\end{align}    
\begin{align}\nonumber
(\mat{D}_{23})_{ij} =
\begin{cases}
	\nu_{23} & (i,j) = (S(I),1) \\
	\rho_2 & (i,j) = (S(I), 3) \\
	\rho_3 & (i,j) \in\{ (S(I)-1, 1),(S(I)-2, 1) \}\\
	0 & \text{otherwise}
\end{cases} 
\end{align}    
\begin{align}\nonumber
(\mat{D}_{34})_{ij} =
\begin{cases}
	\nu_{33} & (i,j) = (S'(I),1) \\
	\rho_3 & (i,j) \in\{ (S'(I),3),(S'(I)-1,1),\\ 
	& \hspace{1.4cm}(S'(I)-2,1)\} \\
	0 & \text{otherwise}
\end{cases} .
 \end{align}

 \subsubsection{Construction of $\mat{B}^{(I+1)}$} 

The matrix $\mat{B}$ describes identical motion of all symmetrically equivalent beads. Now, one can split $\mathcal{T}_1$ into two $\mathcal{\tilde L}_1$-leaves, one $\mathcal{L}_1$- and one $\mathcal{L}_2$-leaf as well as the core bead. Therefore, its structure reads
\begin{align}
\mat{B}^{(I+1)}=
\begin{pmatrix}
	\bm{\upeta} && 0 && \mat{G}_{13} && \mat{G}_{14} \\
	0 && \bm{\updelta} && \mat{G}_{23} && 0 \\
	\mat{G}_{31} && \mat{G}_{32} && \bm{\upepsilon} && \mat{G}_{34} \\
	\mat{G}_{41} && 0 && \mat{G}_{43} &&  \mu_{333}^{(3)}
\end{pmatrix},
\end{align}
where $\bm{\upeta}$ represents a co-phase motion of two $\mathcal{\tilde L}_1$-leaves. The only difference to $\tmat{A}^{(I)}$ being in one entry,
\begin{align}
(\bm{\upeta})_{ij}=
\begin{cases}
	\mu_{133}^{(3)} + \rho_3 & (i,j) = (\widetilde{S}(I), \widetilde{S}(I)) \\
	(\tmat{A}^{(I)})_{ij} & \text{otherwise}
\end{cases}
\end{align}
The other diagonal blocks are given in Eqs.~\eqref{matdelta} and \eqref{matepsilon}. The off-diagonal blocks are as follows,
\begin{align}\nonumber
(\mat{G}_{13})_{ij} &=
\begin{cases}
	\rho_3 & (i,j) = (\widetilde{S}(I), S'(I)) \\
	0 & \text{otherwise}
\end{cases} 
 \end{align}    
\begin{align}\nonumber
(\mat{G}_{14})_{ij} &=
\begin{cases}
	\nu_{33} & (i,j) = (\widetilde{S}(I),1) \\
	\rho_3 & (i,j) \in\{ (\widetilde{S}(I)-1, 1),(\widetilde{S}(I)-2, 1)\} \\
	0 & \text{otherwise}
\end{cases} 
  \end{align}    
\begin{align}\nonumber
(\mat{G}_{23})_{ij} &=
\begin{cases}
	\nu_{23} & (i,j) = (S(I),1) \\
	\rho_2 & (i,j) = (S(I), 3) \\
	\rho_3 & (i,j) \in\{ (S(I)-1, 1),(S(I)-2, 1)\} \\
	0 & \text{otherwise}
\end{cases} 
 \end{align}    
\begin{align}\nonumber
(\mat{G}_{34})_{ij} &=
\begin{cases}
	\nu_{33} & (i,j) = (S'(I),1) \\
	\rho_3 & (i,j) \in\{ (S'(I)-1,1), (S'(I)-2,1)\} \\
	0 & \text{otherwise}
\end{cases} 
 \end{align}    
Furthermore, $\mat{G}_{31}=2\mat{G}_{13}^{\mathrm{T}}$, $\mat{G}_{41}=2\mat{G}_{14}^{\mathrm{T}}$, $\mat{G}_{32}=\mat{G}_{23}^{\mathrm{T}}$, and $\mat{G}_{43}=\mat{G}_{34}^{\mathrm{T}}$.

\subsection{Tree $\mathcal{T}_2$}

\subsubsection{Number of distinct amplitudes}

\begin{figure}[t]
\centering
\includegraphics[width=1\linewidth,trim=0 0 0 0]{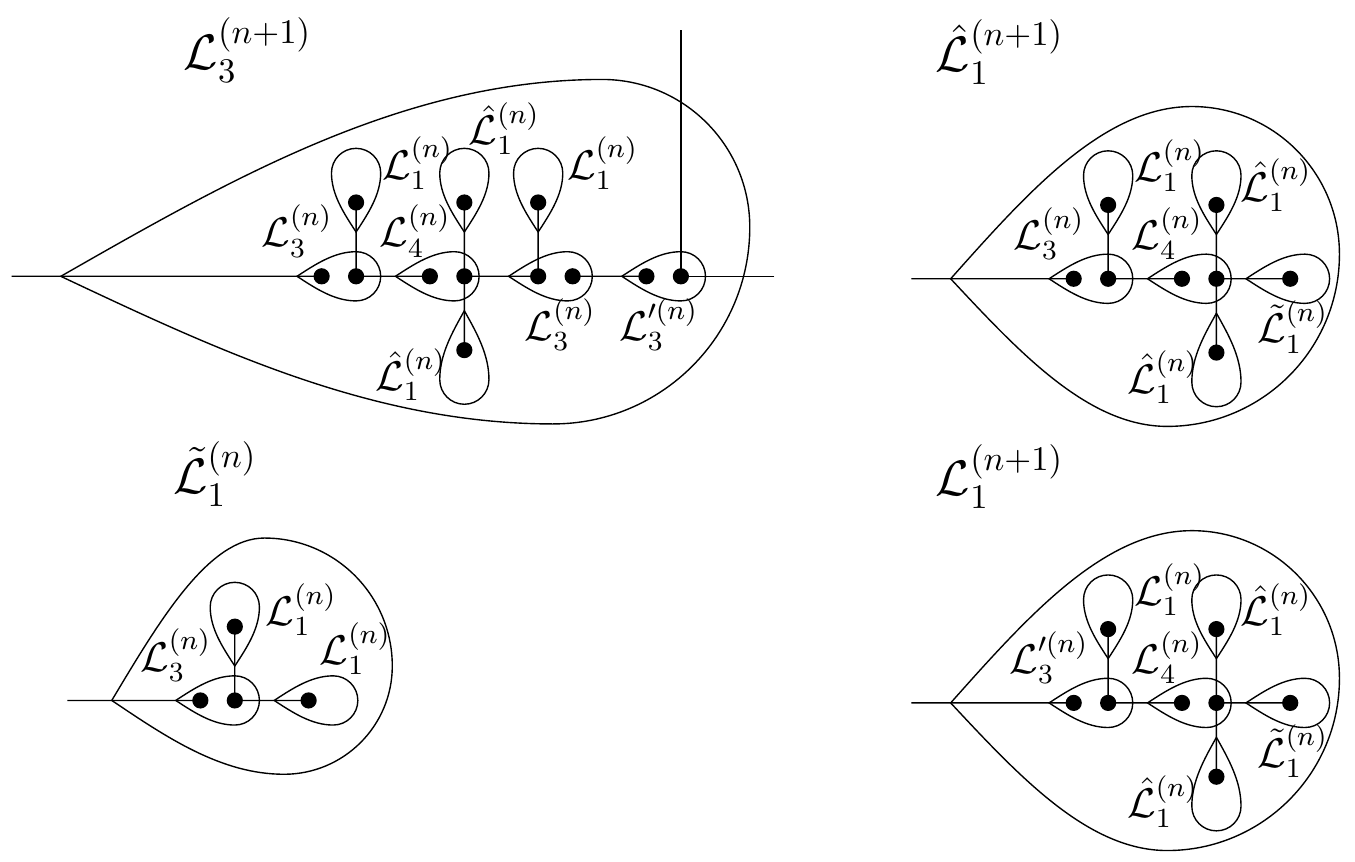}
\caption{Schematic sketch of substructures ("leaves" $\mathcal{L}$) of tree $\mathcal{T}_2$.}\label{leaves_p}
\end{figure}

As for tree $\mathcal{T}_1$, for $\mathcal{T}_2$ the number of distinct amplitudes in a given mode (that is then equal to the size of the corresponding reduced matrices), can be calculated by observation of the iterative construction of leaves (Fig.~\ref{leaves_p}). 

The tree $\mathcal{T}_2$ at iteration $I=n$ consists of two leaves $\mathcal{\tilde {L}}_1^{(n)}$, two leaves $\mathcal{\hat{L}}_1^{(n)}$, and the core connecting the four leaves. All other leaves are substructures of these leaves, e.g., $\mathcal{\tilde {L}}_1^{(n)}$ consists of one leaf $\mathcal{L}_3^{(n)}$ and two leaves $\mathcal{L}_1^{(n)}$, where the leaves $\mathcal{L}_1^{(n)}$ are symmetrically equivalent. With this, denoting by $\widetilde{S}(n)$, $S(n)$, and $S'(n)$ the number of distinct amplitudes in $\mathcal{\tilde {L}}_1^{(n)}$,  $\mathcal{L}_1^{(n)}$, and $\mathcal{L}_3^{(n)}$, respectively, we get
\begin{align}\label{equ:Z(n)}
\widetilde{S}(n)=S(n)+S'(n).
\end{align}
Inspecting leaves $\mathcal{\hat{L}}_1^{(n)}$ and $\mathcal{L}_1^{(n)}$ one finds that $\hat{S}(n)$ (related to $\mathcal{\hat{L}}_1^{(n)}$) is $\hat{S}(n)=S(n)$ and that
\begin{align}
S(n) = 2S(n-1)+2S'(n-1)+\tilde{S}(n-1).
\end{align}
Using Eq.~(\ref{equ:Z(n)}) one gets readily
\begin{align} \label{F_RG_1}
S(n) = 3S'(n-1)+3S(n-1).
\end{align}
Analogously, the structure of leaf $\mathcal{L}_3^{(n)}$, see Fig.~\ref{leaves_p}, yields
\begin{align} \label{F_B_1}
S'(n) = 4S'(n-1)+3S(n-1).
\end{align}
The set of Eqs. (\ref{F_RG_1}) und (\ref{F_B_1}) can be solved under initial conditions $S(1)=S'(1)=1$, $S(2)=6$, and $S'(2)=7$, leading for $\mathcal{T}_2$ to the corresponding line in Eq.~\eqref{Sn} of the main part and to 
\begin{align}\label{Sprime}
S'(n)&=\frac{1}{\sqrt{37}} \left(\frac{7+\sqrt{37}}{2}\right)^n - \frac{1}{\sqrt{37}}\left(\frac{7-\sqrt{37}}{2}\right)^n.
\end{align} 
Using then Eq.~\eqref{equ:Z(n)},  Eq.~\eqref{tildeSm} of the main part for tree $\mathcal{T}_2$ follows. Finally, Eq.~\eqref{S_B} for $\mathcal{T}_2$ reflects the fact that tree $\mathcal{T}_2$ consists of the core and connected to it two symmetrically equivalent pairs of leaves $\mathcal{\tilde {L}}_1$ and $\mathcal{\hat{L}}_1^{(n)}$.

\subsubsection{Initial matrices}

The iterative algorithm of construction of the reduced matrices is initialized by matrix $\mat{A}^{(2)}$ related to $\mathcal{L}_1^{(2)}$-leaf and auxiliary matrix $\mat{H}^{(2)}$,
\begin{align}\nonumber
\mat{A}^{(2)}=&
\left(
\begin{array}{cccccc}
 \mu^{(1)}_{3}+\rho_{3} & \nu_{13} & 0 & \rho_{3} & 0 & 0 \\
 2\nu_{13} & \mu^{(3)}_{114} & 2 \rho_{4} & \nu_{34} & 0 & \rho_{4} \\
 0 & \rho_{4} & \mu^{(1)}_{4}+\rho_{4} & \nu_{14} & 0 & \rho_{4} \\
 2 \rho_{3} & \nu_{34} & 2 \nu_{14} & \mu^{(4)}_{1133} &\rho_{3} &\nu_{34} \\
 0 & 0 & 0 & \rho_{3} & \mu^{(1)}_{3} & \nu_{13} \\
 0 & \rho_{4} & 2 \rho_{4} & \nu_{34} & \nu_{13} & \mu^{(3)}_{134}-\rho_{3} \\
\end{array}
\right)
\end{align}
and
\begin{align}\nonumber
\mat{H}^{(2)}=&
\left(
\begin{array}{ccccccc}
 \mu^{(3)}_{333} & \rho_{3} & \nu_{33} & 0 & \rho_{3} & 0 & 0 \\
 \rho_{3} & \mu^{(1)}_{3} & \nu_{13} & 0 & \rho_{3} & 0 & 0 \\
 \nu_{33} & \nu_{13} & \mu^{(3)}_{134} & 2 \rho_{4} &\nu_{34} & 0 &\rho_{4} \\
 0 & 0 & \rho_{4} &\mu^{(1)}_{4}+\rho_{4} & \nu_{14} & 0 & \rho_{4} \\
 \rho_{3} & \rho_{3} & \nu_{34} & 2 \nu_{14} & \mu^{(4)}_{1133} & \rho_{3} & \nu_{34} \\
 0 & 0 & 0 & 0 & \rho_{3} & \mu^{(1)}_{3} & \nu_{13} \\
 0 & 0 & \rho_{4} & 2 \rho_{4} & \nu_{34} & \nu_{13} & \mu^{(3)}_{144}-\rho_{4} \\
\end{array}
\right).
\end{align}
In general, $\mat{H}^{(I)}$ describes two leaves $\mathcal{L}_3^{(I)}$, each inside two $\mathcal{\tilde L}_1^{(I)}$-leaves that are moving in antiphase, hence the size of $\mat{H}^{(I)}$ is given by $S'(I)$ of Eq.~\eqref{Sprime}.

\subsubsection{Construction of $\mat{\hat{A}}^{(I)}$}

Observing that leaf $\mathcal{\hat{L}}_1^{(I)}$ differs from $\mathcal{L}_1^{(I)}$ only in the functionality of the bead that is connected to these leaves, $\mat{\hat{A}}^{(I)}$ can be easily obtained by changing only the last element of $\mat{A}^{(I)}$
\begin{align}
\mat{\hat{A}}^{(I)}=
\begin{cases}
\mu^{(3)}_{144}-\rho_{4} \qquad &\text{for } (i,j)=(S(I),S(I)) \\
\left(\mat{A}^{(I)}\right)_{ij} \qquad &\text{else}
\end{cases}
\end{align}

\subsubsection{Construction of $\mat{\tilde{A}}^{(I)}$}

Matrix $\mat{\tilde{A}}^{(I)}$ is related to antiphase motions of two leaves $\mathcal{\tilde {L}}_1^{(I)}$. It has the following form
\begin{align}
\mat{\tilde{A}}^{(I)}=
\begin{pmatrix}
\bm{\upalpha} & \mat{C_{12}} \\
\mat{C_{21}} & \mat{H}^{(I)} \\
\end{pmatrix}
\end{align}
The elements of the matrices $\mat{C_{12}}$ and $\mat{C_{21}}$ reflect the connection of leaves $\mathcal{L}_3^{(I)}$ and $\mathcal{L}_1^{(I)}$ inside $\mathcal{\tilde {L}}_1^{(I)}$. Matrix $\bm{\upalpha}$ describes a co-phase motion of two $\mathcal{L}_1^{(I)}$-leaves inside $\mathcal{\tilde {L}}_1^{(I)}$, therefore it can be readily obtained from $\mat{A}^{(I)}$ describing an antiphase of these leaves, by replacing the element related to the bead lying at the edge of the leaves. Explicitly this means
\begin{align}
\left(\bm{\upalpha}\right)_{ij}=
\begin{cases}
\mu^{(3)}_{134}+\rho_{3} \qquad &\text{for } (i,j)=(S(I),S(I))\\
\left(\mat{A}^{(I)}\right)_{ij} \qquad &\text{else}
\end{cases}
\end{align}
The connection matrix $\mat{C_{12}}$ reads
\begin{align}\nonumber
\left(\mat{C_{12}}\right)_{ij}=
\begin{cases}
\nu_{33} \qquad &\text{for } (i,j)=(S(I),1) \\
\rho_{3} \qquad &\text{for } (i,j)\in\{S(I)-1,1),(S(I)-2,1),\\ 
	&\hspace{1.4cm} (S(I),3)\} \\
0 \qquad &\text{else}
\end{cases}
\end{align}
Due to the symmetry of the dynamical matrix $\mat{C_{21}}=2\mat{C_{12}}^{\mathrm{T}}$ holds, where $^{\mathrm{T}}$ denotes transposition.

\subsubsection{Construction of $\mat{A}^{(I+1)}$ and $\mat{H}^{(I+1)}$}

For the construction of $\mat{A}^{(I+1)}$ and $\mat{H}^{(I+1)}$ it is convenient to introduce another auxiliary matrix $\mat{\tilde{H}}^{(I)}$: 
\begin{align}
\mat{\tilde{H}}^{(I)}=
\begin{pmatrix}
\bm{\upgamma} & \mat{D_{12}} & \mat{D_{13}} \\
\mat{D_{21}} & \bm{\upbeta} & \mat{D_{23}} \\
\mat{D_{31}} & \mat{D_{32}} & \bm{\upgamma} \\
\end{pmatrix},
\end{align}
where
\begin{align}
\bm{\upbeta}=
\begin{cases}
\mu^{(3)}_{134} \qquad &\text{for } (i,j)=(S(I), S(I)) \\
\left(\mat{A}^{(I)}\right)_{ij} \qquad &\text{else},
\end{cases}
\end{align}
\begin{align}
\bm{\upgamma}=
\begin{cases}
\mu^{(3)}_{134} \qquad &\text{for } (i,j)=(S'(I),S'(I)) \\
\left(\mat{H}^{(I)}\right)_{ij} \qquad &\text{else},
\end{cases}
\end{align}
\begin{align}\nonumber
\mat{D_{12}}=\mat{D_{21}}^{\mathrm{T}}&=
\begin{cases}
\rho_{3} \qquad &\text{for } (i,j)=(S'(I),S(I)) \\
0 \qquad &\text{else},
\end{cases}
\end{align}    
\begin{align}\nonumber
\mat{D_{13}}=\mat{D_{31}}^{\mathrm{T}}&=
\begin{cases}
\nu_{33} \qquad &\text{for } (i,j)=(S'(I),1) \\
\rho_{3} \qquad &\text{for } (i,j) \in \{(S'(I),3),(S'(I)-1,1),\\ 
	&\hspace{1.4cm}  (S'(I)-2,1)\} \\
0 \qquad &\text{else},
\end{cases}
\end{align}
and    
\begin{align}\nonumber
\mat{D_{23}}=\mat{D_{32}}^{\mathrm{T}}&=
\begin{cases}
\nu_{33} \qquad &\text{for } (i,j)=(S(I),1) \\
\rho_{3} \qquad &\text{for } (i,j) \in \{(S(I),3),(S(I)-1,1),\\ 
	&\hspace{1.4cm}  (S(I)-2,1)\} \\
0 \qquad &\text{else}.
\end{cases}
\end{align}

Now it is possible to construct the $\mat{A}^{(I+1)}$ related to the whole $\mathcal{L}_1^{(I+1)}$-leaf (see Fig.~\ref{leaves_p}).
\begin{align}
\mat{A}^{(I+1)}=
\begin{pmatrix}
 \bm{\updelta} & \mat{E_{12}} & \mat{E_{13}} \\
\mat{E_{21}} & \bm{\upepsilon} & \mat{E_{23}} \\
\mat{E_{31}} & \mat{E_{32}}&\bm{\upzeta} \\
\end{pmatrix}
\end{align}
where
\begin{align}
\bm{\updelta}&=
\begin{cases}
\mu^{(3)}_{144} \qquad &\text{for } (i,j)=(\tilde{S}(I),\tilde{S}(I)) \\
\left(\mat{\tilde{A}}^{(I)}\right)_{ij} \qquad &\text{else}
\end{cases} 
\end{align} 
is related to $\mathcal{\tilde L}_1^{(I)}$ inside $\mathcal{L}_1^{(I+1)}$,
\begin{align}
\bm{\upepsilon} &=
\begin{cases}
\mu^{(3)}_{144}+ \rho_{4} \qquad &\text{for } (i,j)=(S(I),S(I)) \\
\left(\mat{A}^{(I)}\right)_{ij} \qquad &\text{else}
\end{cases}
\end{align}  
 to two co-phasely moving symmetrically equivalent $\mathcal{\hat L}_1^{(I)}$ inside $\mathcal{L}_1^{(I+1)}$, and 
\begin{align}
\bm{\upzeta}&=
\begin{cases}
\mu^{(3)}_{134}-\rho_{3} \qquad &\text{for } (i,j)=(2S'(I)+S(I),2S'(I)+S(I)) \\
\mu^{(4)}_{3333}  \qquad &\text{for } (i,j)=(1,1) \\
\mu^{(3)}_{144}  \qquad &\text{for } (i,j)=(3,3) \\
\nu_{34}  \qquad &\text{for } (i,j) \in\{(3,1),(1,3)\} \\
\left(\mat{\tilde{H}}^{(I)}\right)_{ij} \qquad &\text{else}
\end{cases}
\end{align}
to the leaves $\mathcal{L}_4^{(I)}$, $\mathcal{L'}_3^{(I)}$, and $\mathcal{L}_1^{(I)}$ inside $\mathcal{L}_1^{(I+1)}$.
The connection blocks are
\begin{align}\nonumber
\mat{E_{12}}=2\mat{E_{21}}^{\mathrm{T}}&=
\begin{cases}
2 \rho_{4} \qquad &\text{for } (i,j)=(\tilde{S}(I),S(I)) \\
0 \qquad &\text{else},
\end{cases}
\end{align}    
\begin{align}\nonumber
\mat{E_{13}}=\mat{E_{31}}^{\mathrm{T}}&=
\begin{cases}
\nu_{34} \qquad &\text{for } (i,j)=(\tilde{S}(I),1) \\
\rho_{4} \qquad &\text{for } (i,j)=(\tilde{S}(I),3) \\
\rho_{3} \qquad &\text{for } (i,j) \in \{ (\tilde{S}(I)-1,1), \\ 
	&\hspace{1.4cm}(\tilde{S}(I)-2,1)\} \\
0 \qquad &\text{else},
\end{cases}
\end{align} 
and   
\begin{align}\nonumber
2\mat{E_{23}}=\mat{E_{32}}^{\mathrm{T}}&=
\begin{cases}
2\nu_{34} \qquad &\text{for } (i,j)=(S(I),1) \\
2\rho_{4} \qquad &\text{for } (i,j)=(S(I),3) \\
2\rho_{3} \qquad &\text{for } (i,j) \in \{ (S(I)-1,1), \\ 
	&\hspace{1.4cm} (S(I)-2,1)\} \\
0 \qquad &\text{else}
\end{cases}
\end{align}

The auxiliary matrix $\mat{H}^{(I+1)}$ can be constructed from already known parts. One gets
\begin{align}
\mat{H}^{(I+1)} =
\begin{cases}
\mu^{(3)}_{144}-\rho_{4} \qquad &\text{for } (i,j)=(S'(I+1),\\ 
	&\hspace{1.4cm} S'(I+1)) \\
\mu^{(3)}_{333} \qquad &\text{for } (i,j)=(1,1) \\
\mu^{(3)}_{134} \qquad &\text{for } (i,j)=(3,3) \\
\nu_{33} \qquad &\text{for } (i,j)\in\{(1,3),(3,1)\} \\
\left(\mat{H}^{(I+1)}_1\right)_{ij} \qquad &\text{else},
\end{cases}
\end{align}
where
\begin{align}
\mat{H}^{(I+1)}_1=
\begin{pmatrix}
\bm{\upeta} & \mat{F_{12}} & \mat{F_{13}} \\
\mat{F_{21}} & \bm{\upepsilon} & \mat{F_{23}} \\
\mat{F_{31}} & \mat{F_{32}}& \bm{\upeta} \\
\end{pmatrix}
\end{align}
with
\begin{align}
\bm{\upeta}&=
\begin{cases}
\mu^{(3)}_{144} \qquad &\text{for } (i,j)=(2S'(I)+S(I),2S'(I)+S(I)) \\
\left(\bm{\upzeta}\right)_{ij} \qquad &\text{else}
\end{cases}
\end{align}
and
\begin{align}\nonumber
\mat{F_{12}}=2\mat{F_{21}}^{\mathrm{T}}&=
\begin{cases}
2 \rho_{4} \qquad &\text{for } (i,j)=(2S'(I)+S(I),S(I)) \\
0 \qquad &\text{else}
\end{cases}
\end{align}    
\begin{align}\nonumber
\mat{F_{13}}=\mat{F_{31}}^{\mathrm{T}}&=
\begin{cases}
\nu_{34} \qquad &\text{for } (i,j)=(2S'(I)+S(I),1) \\
\rho_{4} \qquad &\text{for } (i,j)=(2S'(I)+S(I),3) \\
\rho_{3} \qquad &\text{for } (i,j)\in \{(2S'(I)+S(I)-1,1),\\ 
	&\hspace{1.4cm} (2S'(I)+S(I)-2,1)\} \\
0 \qquad &\text{else}
\end{cases}
\end{align}    
\begin{align}\nonumber
2\mat{F_{23}}=\mat{F_{32}}^{\mathrm{T}}&=
\begin{cases}
2\nu_{34} \qquad &\text{for } (i,j)=(S(I),1) \\
2\rho_{4} \qquad &\text{for } (i,j)=(S(I),3) \\
2\rho_{3} \qquad &\text{for } (i,j)\in \{(S(I)-1,1),\\ 
	&\hspace{1.4cm} (S(I)-2,1)\} \\
0 \qquad &\text{else}.
\end{cases}
\end{align}

\subsubsection{Construction of $\mat{B}^{(I)}$}

The reduced matrices related to the modes in which the core of the tree is mobile read
\begin{align}
\mat{B}^{(I)} &=
\begin{pmatrix}
\bm{\uptheta} & \mat{I_{12}} & \mat{I_{13}} \\
\mat{I_{21}} & \bm{\upepsilon} & \mat{I_{23}} \\
\mat{I_{31}} &\mat{I_{32}} & \mu^{(4)}_{3333}
\end{pmatrix}
\end{align}    
\begin{align}
\bm{\uptheta} &=
\begin{cases}
\mu^{(3)}_{144}+\rho_{4} \qquad &\text{for } (i,j)=(\tilde{S}(I), \tilde{S}(I)) \\
\left(\mat{\tilde{A}}^{(I)}\right)_{ij} \qquad &\text{else}
\end{cases}
\end{align}    
\begin{align}\nonumber
\mat{I_{12}}=\mat{I_{21}}^{\mathrm{T}} &=
\begin{cases}
2\rho_{4} \qquad &\text{for } (i,j)=(\tilde{S}(I), S(I)) \\
0 \qquad &\text{else}
\end{cases}
\end{align}    
\begin{align}\nonumber
2\mat{I_{13}}=\mat{I_{31}}^{\mathrm{T}} &=
\begin{cases}
2\nu_{34} \qquad &\text{for } (i,j)=(\tilde{S}(I), 1) \\
2\rho_{3} \qquad &\text{for } (i,j)\in\{(\tilde{S}(I)-1,1),\\ 
	&\hspace{1.4cm} (\tilde{S}(I)-2, 1)\} \\
0 \qquad &\text{else}
\end{cases}
\end{align}    
\begin{align}\nonumber
2\mat{I_{23}}=\mat{I_{32}}^{\mathrm{T}} &=
\begin{cases}
2\nu_{34} \qquad &\text{for } (i,j)=(S(I), 1) \\
2\rho_{3} \qquad &\text{for } (i,j)\in\{(S(I)-1, 1),\\ 
	&\hspace{1.4cm} (S(I)-2, 1)\} \\
0 \qquad &\text{else}
\end{cases}
\end{align}


%

\end{document}